\documentclass[twocolumn,preprintnumbers,amsmath,amssymb,aps,pre,reprint,longbibliography]{revtex4-2}
\usepackage{graphicx}
\usepackage{xcolor}
\begin{document}

\title{Ratchet Effects in Cyclic Pattern Formation Systems with Competing Interactions}
\author{
C. Reichhardt and C. J. O. Reichhardt 
} 
\affiliation{
Theoretical Division and Center for Nonlinear Studies,
Los Alamos National Laboratory, Los Alamos, New Mexico 87545, USA
}

\date{\today}

\begin{abstract}
Ratchet effects can appear for particles interacting with an asymmetric potential under ac driving or for a thermal system in which a substrate is periodically flashed. Here, we show that a new type of collective ratchet effect can arise for a pattern-forming system coupled to an asymmetric substrate when the interaction potential between the particles is periodically oscillated in order to cycle the system through different patterns. We consider particles with competing short-range attraction and long-range repulsion subjected to time-dependent oscillations of the ratio between the attractive and repulsive interaction terms, which causes the system to cycle periodically between crystal and bubble states. In the presence of the substrate, this system exhibits both a positive and a reversed ratchet effect, and we show that there is a maximum in the ratchet efficiency as a function of interaction strength, ac drive frequency, and particle density. Our results could be realized for a variety of pattern-forming systems on asymmetric substrates where the pattern type or particle interactions can be oscillated.
\end{abstract}

\maketitle

\section{Introduction}

In a ratchet effect, an ac drive produces a net dc drift of particles
if enough symmetries in the system are broken, such as by coupling the
particles to an asymmetric substrate \cite{Magnasco93,Faucheux95,Reimann02}.
When the ac drive originates from
an externally applied drive, a rocking ratchet can appear,
and if thermal fluctuations are present,
a flashing ratchet effect can be produced by turning
the substrate on and off
periodically \cite{Astumian02,Rousselet94}.
Ratchet effects can occur at the single-particle level, which generally leads
to a net flux in the easy direction of the substrate asymmetry.
In systems of multiple interacting particles,
non-monotonic ratchet efficiencies as well as reversals of the
ratchet direction can occur when collective effects become important
\cite{deSouzaSilva06a,Dinis07,Lu07}.
Ratchet effects have been studied for a wide variety of systems,
including soft matter systems such as colloidal
particles \cite{Rousselet94,Marquet02},
granular matter \cite{Farkas02,Wambaugh02},
or
active matter \cite{Reichhardt17a},
superconducting vortices on asymmetric substrates \cite{Lee99,Wambaugh99,Yu07},
biological systems \cite{Lau17},
Wigner crystals \cite{Reichhardt23a},
magnetic skyrmions \cite{Reichhardt15a},
dusty plasmas \cite{He20},
cold atoms on optical substrates \cite{MenneratRobilliard99,Salger09},
and quantum systems \cite{Linke99,Grossert16}. 

Ratchet effects have been extensively studied for point-like particles, but there are also some examples of ratchet effects for
polymer-like chains \cite{Kauttonen10}.
In certain magnetic \cite{Chen19}
and liquid crystal \cite{Ackerman17}
skyrmion systems, where the objects are bubble-like
instead of particle-like, a ratchet effect can arise
due to periodic oscillation of the object shape
even in the absence of a substrate.
There is also work
on Leidenfrost ratchets \cite{Linke06},
where a fluid bubble interacting with a heated, asymmetric
periodic substrate can undergo directed motion.
In general, relatively little is known about ratchet effects
for extended or flexible objects. When systems of this type
are coupled to an asymmetric substrate,
it may be possible to generate new kinds of ratchet effects
by periodically
changing the size or shape of the objects or
the interactions between the particles out of which
each larger-scale object is composed
without needing to flash the substrate or apply an ac shaking drive.

Particles with competing short-range attractive
and long range repulsive (SALR) interactions are known
to form complex patterns of extended objects, including
crystals, stripes, labyrinths, bubbles, and void lattices
\cite{Seul95,Stoycheva00,Reichhardt03,Reichhardt04,Mossa04,Sciortino04,Nelissen05,Liu08,Reichhardt10,McDermott14,Liu19,AlHarraq22,Hooshanginejad24}.
Similar patterns appear for particles with purely repulsive interactions
when the interaction potential contains two steps or has at least
two distinct length scales
\cite{Jagla98,Malescio03,Glaser07}.
Interactions of this type can arise in soft matter and biological
systems
\cite{Malescio03,Glaser07,CostaCampos13,AlHarraq22,Hooshanginejad24} as
well as 
in hard condensed matter systems,
where
bubbles and stripes
are predicted to form in electron liquid crystals
\cite{Fogler96,Moessner96,Cooper99,Fradkin99,Gores07,Zhu09,Friess18}.
Additionally, in
multiple component superconductors
the vortex-vortex interactions can have
multiple length scales, giving rise to mesoscale stripe and bubble
ordering \cite{Xu11,Komendova12,Varney13,Sellin13,Brems22}.
Bubble like patterns
can also occur for magnetic textures such as
skyrmions \cite{Reichhardt22a} as well as in liquid crystal 
systems \cite{Sohn18}. 

If a pattern-forming system with competing interactions
is held at a fixed particle density and
the short-range attraction is set to zero,
a uniform crystal appears.
As the attraction is increased, the system transitions into a 
stripe state and then
forms a bubble lattice \cite{Reichhardt10,Nelissen05,Liu08,McDermott16}. 
When a periodic quasi-one-dimensional
asymmetric substrate is added to the system and the
particles are subjected to ac driving, ratcheting motion appears in which
the efficiency of the ratcheting depends on the type of pattern that
is present.
A strong ratchet effect occurs in the stripe phase when 
the stripes are able to align with
the substrate troughs \cite{Reichhardt24a}.
In the bubble regime, when the bubbles are large and extend over
several substrate minima, a weak ratchet effect is present, but when the
size of the bubbles is reduced enough to allow each bubble to fit inside
a single substrate minimum, the ratchet effect becomes strong again.

In this work, we propose that a new type of ratchet effect can occur
on an asymmetric substrate without application of ac driving or flashing
of the substrate. Instead,
the interaction between the particles oscillates
so that the system periodically switches between
crystal, stripe, and bubble states as a function of time.
We find that this system can exhibit a strong ratchet effect
in which individual particles and bubbles can be transported
across the substrate, and we map out the ratchet efficiency as a
function of substrate spacing and strength, particle density,
and pulse frequency.
The ratchet effect arises due to
the asymmetric spreading of the bubbles
that occurs as the attraction is decreased.
For weak substrates, we find a reversed ratchet effect
in which the contraction of the crystal to
a bubble state becomes asymmetric.
Our results provide another way
to realize ratchet effects by changing the particle interactions
and hence the pattern formed by the system,
and could be generalized to the broad class of
bubble and pattern-forming systems coupled to an asymmetric
substrate.
Specific examples of where our results could be applied
include skyrmion bubbles coupled to an asymmetric substrate
under a periodic magnetic field that changes the size of the skyrmions
from large to small, as well as
certain colloidal systems with competing interactions
in which the ratio of repulsion to attraction can be controlled with a
periodically modulated magnetic field \cite{Hooshanginejad24}.    

\section{Simulation}

We consider a two-dimensional system
of size $L \times L$, where $L=36$,
with periodic boundary conditions in the $x$- and $y$-directions.
The system contains $N$ particles of density $\rho=N/L^2$ interacting
with a quasi-one-dimensional
asymmetric periodic substrate.
The SALR interaction potential between
particles has a long-range repulsive Coulomb term
and a short-range attractive term, and is given by:
\begin{equation} V(R_{ij}) = \frac{1}{R_{ij}} - B\exp(-\kappa R_{ij}) \ . \end{equation}
The distance between particles $i$ and $j$ is $R_{ij}=|{\bf R}_i-{\bf R}_j|$.
The first term is the Coulomb repulsion, which favors
formation of a uniform crystal.
The second term is attractive and has strength $B$
and an inverse range of $\kappa$.
In this work, we fix $\kappa=1.0$ and
impose a time variation on $B=B(t)$.
In previous work, particles with this interaction potential were shown
to form crystal, stripe, and bubble phases
as a function of increasing $B$ \cite{Reichhardt10,Reichhardt24a}.

The particle dynamics are obtained using the following overdamped equation:
\begin{equation}
\eta \frac{d {\bf R}_{i}}{dt} = -\sum^{N}_{j \neq i} \nabla V(R_{ij}) + {\bf F}^{s}_{i} + {\bf F}_{\rm ac} , 
\end{equation}
Here, $\eta$ is the damping term, which we set to $\eta=1.0$.
The size of our molecular dynamics time step
is $\delta t = 0.0005$.
The first term on the right is the particle-particle interaction force
from Eq.~(1), while
the second term is the force from the substrate,
${\bf F}^{s}_{i}=\nabla U(x_i)$, where $x_i$ is the $x$ coordinate of particle
$i$.
The substrate potential has the same asymmetric form
used in previous ratchet studies
\cite{Lee99,Reimann02,Reichhardt15a,Reichhardt23a}, 
\begin{equation}
U(x_i) = -U_{p}[\sin(2\pi x_i/a) + 0.25\sin(4\pi x_i/a)] \ ,
\end{equation}
where
$a$ is the substrate lattice constant. In this work, 
we characterize the substrate force by $A_{p} = U_{p}/2\pi$,
so that the force in the easy  or $+x$
direction is $F^{\rm easy}_{p} = 0.725A_{p}$, 
and the force in the hard or $-x$ direction is $F^{\rm hard}_{p} =1.5A_p$.
In previous work, an ac external driving force
was applied, but in this work,
we oscillate the coefficient of the second term in the
particle-particle interaction potential
from Eq.~(1),
$B = B_{\max}|\cos(\omega_1 t)|$, where $B_{\max}$ is the maximum value
of $B$ during the cycle.
Since we are using the absolute value of the cosine,
the system passes through two cycles from $B_{\max}$ to $0$
in one period of the cosine,
so we will describe the behavior in terms of the frequency of a single cycle,
$\omega = 2\omega_1$.
When $B$ is zero, the particles form a
crystal state,
while at large $B$, a bubble state appears.
We measure the ratchet velocity
$\langle V \rangle = \sum_{i=1}^N \mathbf{v}_i \cdot \hat{\mathbf{x}}$ by
averaging over 100 oscillation cycles in terms of $\omega$, which is the
same as 50 cycles in terms of $\omega_1$.
The initial particle configurations are obtained by
placing the particles in a uniform triangular lattice and allowing them to
relax into a patterned states.
Typically, we find that there is a transient time during which the
ratcheting motion stabilizes,
so we wait for 20 to 100 oscillation cycles before
beginning to collect velocity data.

\begin{figure}
\includegraphics[width=\columnwidth]{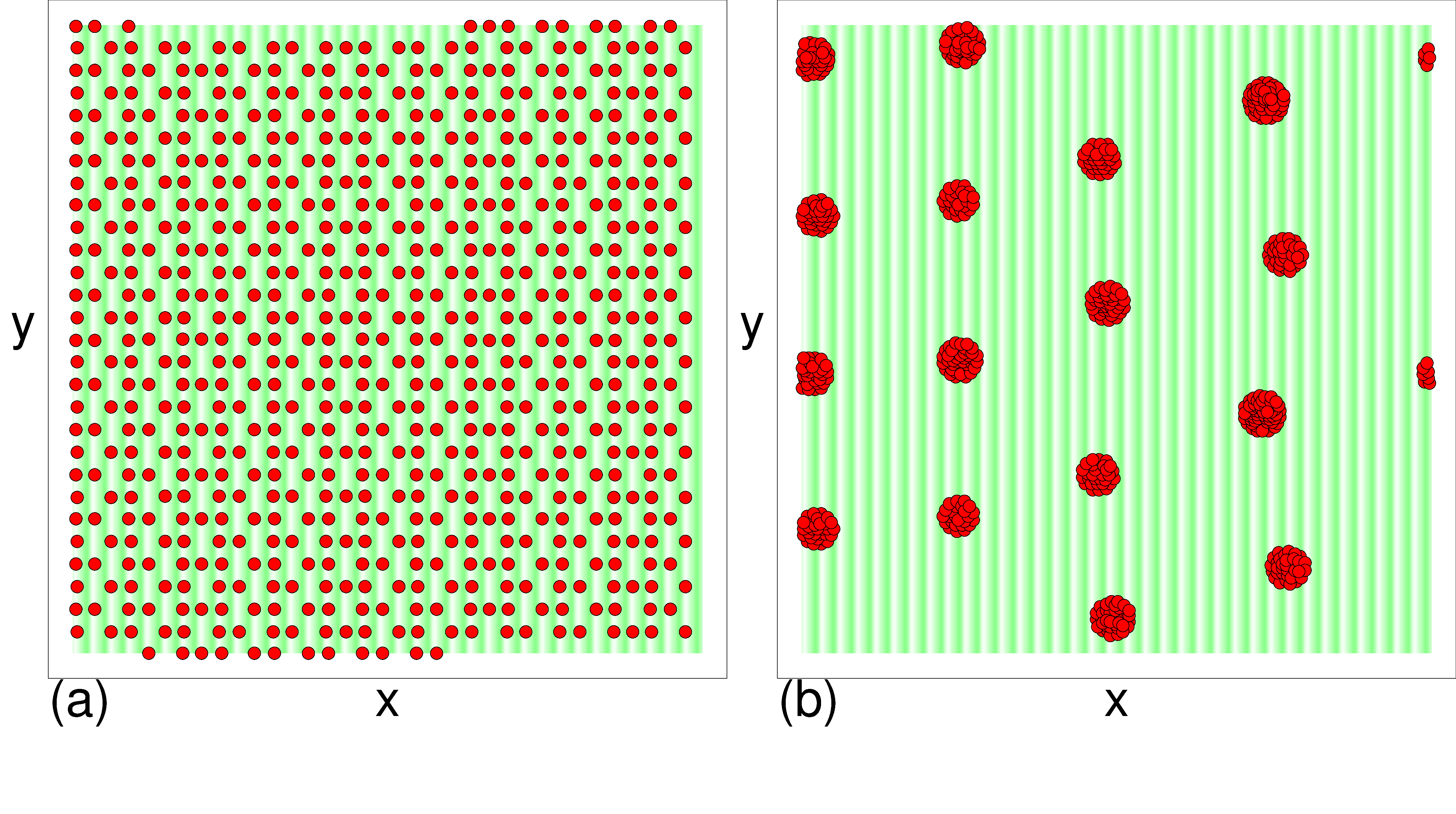}
\caption{Image of particle positions (red) and the underlying
asymmetric potential (green shading, where darker shading is a potential
maximum)  
for a system with $A_p = 2.0$, $\rho = 0.52$,
and $N_p=35$ substrate minima.
(a) A modulated crystal at $B = 0.0$.
(b) The bubble phase at $B = 4.2$.
Throughout this work,
the circles representing the particles are of a size chosen for visualization
purposes only; the actual particles do not overlap in the way that hard
disks would.
} 
\label{fig:1}
\end{figure}

\section{Results}

In Fig.~\ref{fig:1} we illustrate the particle positions and the
asymmetric periodic substrate potential,
where the darker green regions correspond to the potential maxima,
for a system with $A_p = 2.0$, $\rho = 0.52$, and
$N_p=35$ substrate minima.
In this case, we do not oscillate $B$,
but instead show the final configurations
after the particles have been allowed to relax from their initial positions.
At $B=0.0$ in Fig.~\ref{fig:1}(a), there is only Coulomb repulsion between
the particles, so they
spread out to form a uniformly dense crystal.
When $B=4.2$, shown in Fig.~\ref{fig:1}(b), a bubble lattice
appears. For
the parameters we are considering here,
in the absence of a substrate
the system forms an anisotropic crystal when $B < 2.4$
and a bubble state when $B > 2.4$.

\begin{figure}
\includegraphics[width=\columnwidth]{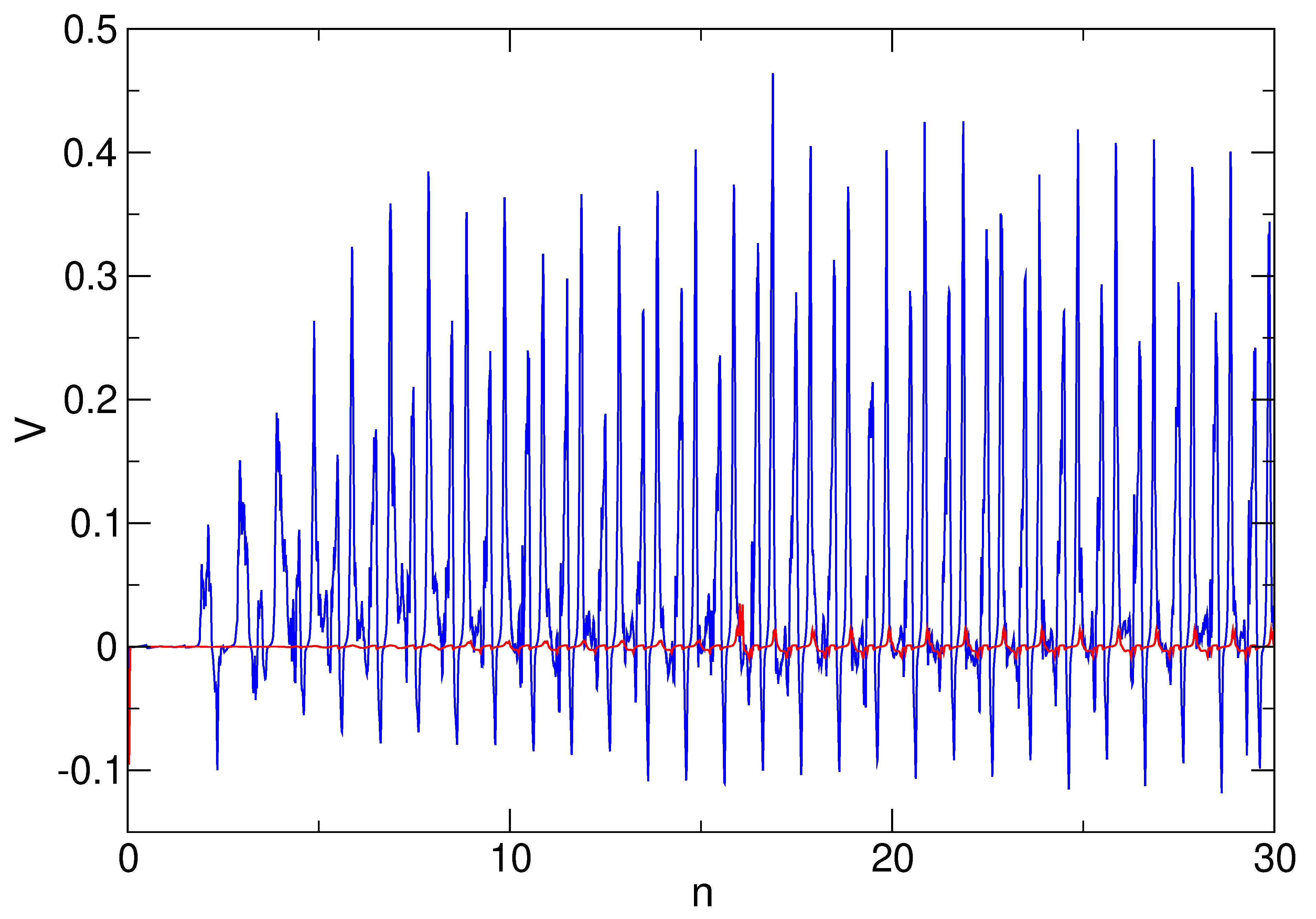}
\caption{The instantaneous average velocity per particle $V$
vs time in cycle numbers $n$
for the system from Fig.~\ref{fig:1} with $A_p = 2.0$,
$N_p=35$, and $\omega/2\pi=2.5 \times 10^{-5}$.
At $\rho = 0.52$ (black) there is
a strong positive velocity response or forward ratchet effect.
At $\rho = 0.0925$ (red),
the time-averaged velocity $\langle V\rangle = 0.0$, indicating
that there is no net ratchet motion.
}
\label{fig:2}
\end{figure}

\begin{figure}
\includegraphics[width=\columnwidth]{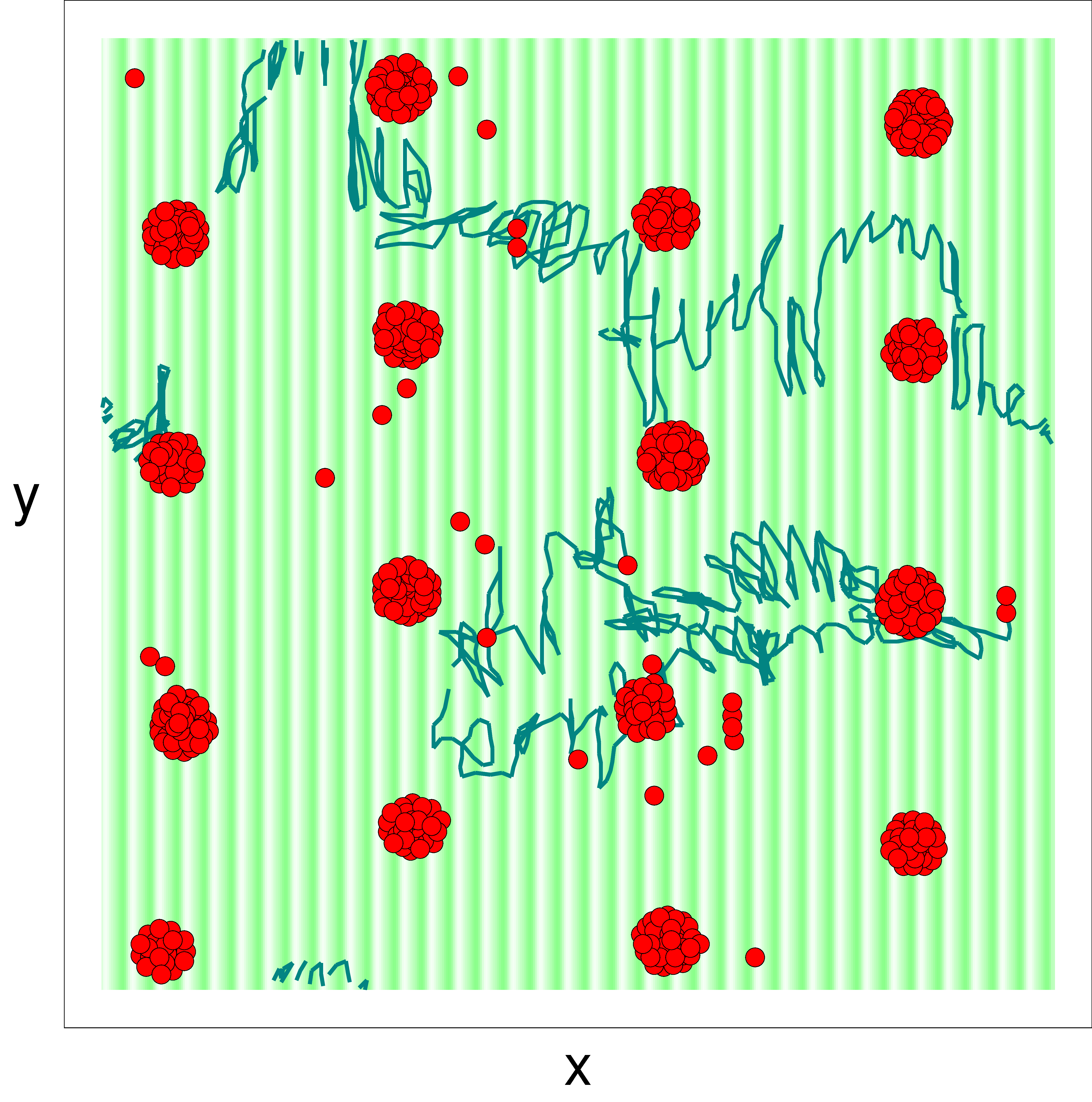}
\caption{Image of particle positions (red) and the underlying
asymmetric potential (green) along with
trajectories (lines) of four particles during 15 oscillation cycles
for the system in Fig.~\ref{fig:2} with $A_p=2.0$,
$N_p=35$, $\omega/2\pi=2.5 \times 10^{-5}$, and
$\rho=0.52$.
The trajectories indicate that there is a net
motion of the particles in the positive $x$ direction.
}
\label{fig:3}
\end{figure}

We next begin periodically oscillating $B$ as a function of time over
the range $B=0$ to $B=4.2$,
so that the system cycles between a uniform crystal and a bubble state.
In Fig.~\ref{fig:2}, we plot the instantaneous average velocity per
particle $V$ versus time in
cycle numbers $n$ for the system from Fig.~\ref{fig:1}
with $A_p=2.0$ and $N_p=35$
at $\omega/2\pi=2.5 \times 10^{-5}$ and two different particle densities.
For $\rho=0.52$,
after a few cycles, the velocity begins to oscillate periodically with
strong positive direction velocity peaks,
indicating that there is a net transport of particles
or a ratchet effect in the positive $x$-direction.
In contrast, at
a lower density of $\rho = 0.0925$, there is no ratchet effect
and the time-averaged velocity $\langle V\rangle=0.0$.
If we set $A_p=0.0$ so that the substrate is absent, we
find $\langle V \rangle = 0.0$ for all particle densities,
since as $B$ is oscillated,
equal numbers of particles move in the positive and negative
$x$ direction during
the expansion and collapse portions of the bubble formation cycle.
In Fig.~\ref{fig:3}, to directly illustrate the translation of the
particles, we show an image of the $\rho=0.52$ system with the trajectories of
four particles during the course of 15 oscillation cycles highlighted.
The motion of the particles is
partially disordered.

\begin{figure}
\includegraphics[width=\columnwidth]{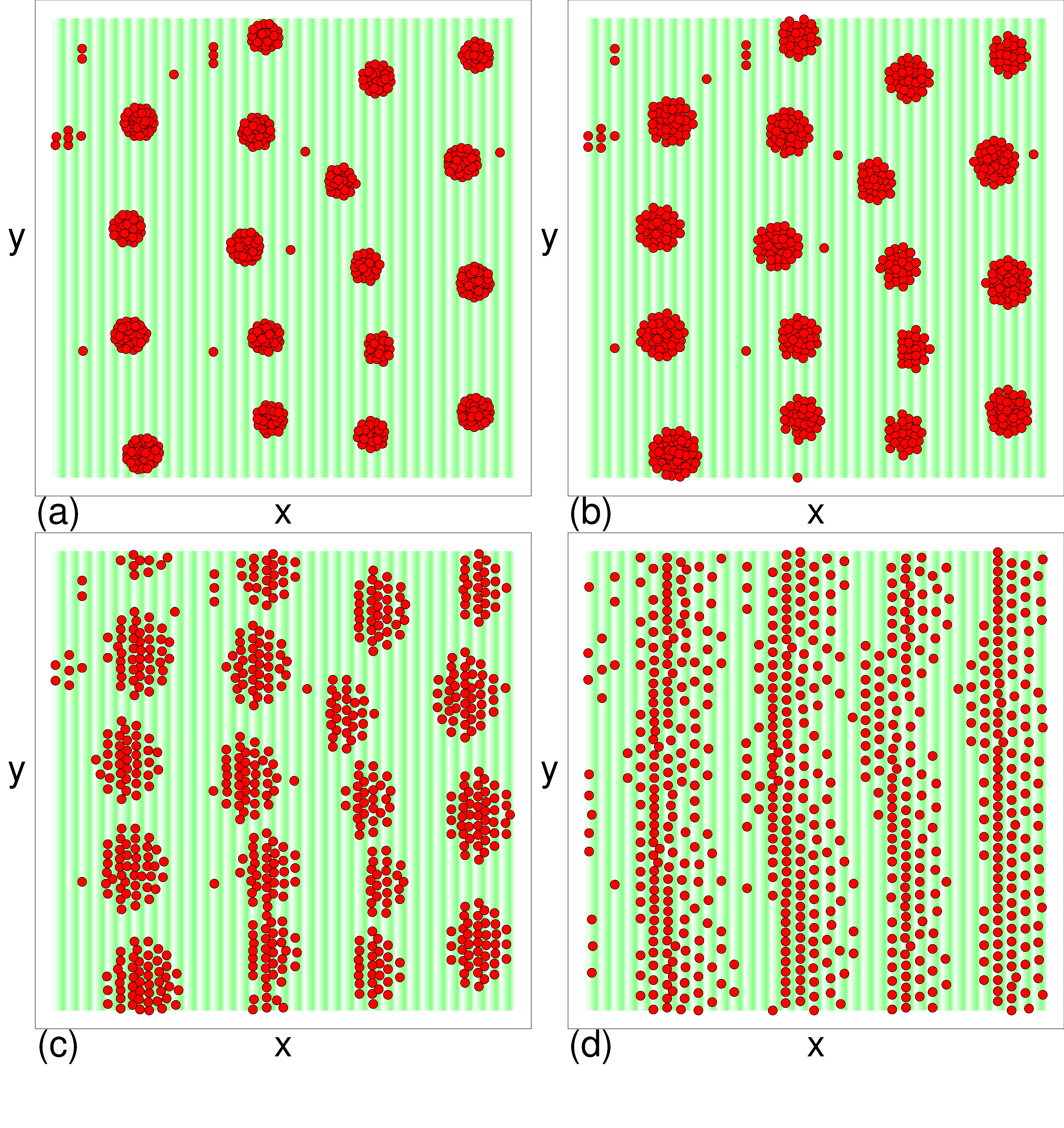}
\caption{Images of particle positions (red) and the substrate
potential (green) taken during the first half of a cycle
with $B$ decreasing from $B = 4.2$ to $B = 0.0$
for the system in Fig.~\ref{fig:3} with $A_p=2.0$, $N_p=35$,
$\omega/2\pi=2.5 \times 10^{-5}$, and $\rho=0.52$.
(a) At $B = 4.2$, the bubbles are compact.
(b) At $B = 2.8$, the bubbles are larger and are expanding.
(c) At $B = 1.8$, the spreading of the bubbles has become
asymmetric due to the substrate.
(d) At $B = 0.0$, the substrate traps the particles and prevents them
from forming a uniform lattice, but due to the asymmetric spreading
process,
there is a density gradient in each
stripe pattern formed by the particles.
}
\label{fig:4}
\end{figure}

To more clearly show how the ratchet effect arises as a result
of oscillating
the strength of the attractive particle interactions,
in Fig.~\ref{fig:4} we show the particle positions at different moments
during the first half of a cycle
for the system in Fig.~\ref{fig:3}
when $B$ is decreasing from $B = 4.2$ to $B = 0.0$.
At the start of the cycle, $B=4.2$ and compact bubbles are present,
as shown in Fig.~\ref{fig:4}(a).
The bubbles begin to expand as $B$ increases, and when $B=2.8$ they have
reached the size illustrated in Fig.~\ref{fig:4}(b).
Once the bubbles become large enough, the expansion becomes asymmetric due
to the underlying substrate, as shown
in Fig.~\ref{fig:4}(c) at $B = 1.8$.
By the time $B$ has reached $B=0.0$, shown in Fig.~\ref{fig:4}(d), the
particles have been trapped by the substrate into a stripe-like
pattern with a clear density gradient in each stripe.
We note that the $B=0.0$ pattern in Fig.~\ref{fig:4}(d), produced by
starting the particles in a bubble state and reducing the attractive
interaction,
differs from the $B = 0.0$ pattern
shown in Fig.~\ref{fig:1}(a),
which is a uniform ground state produced by starting the particles in a
lattice and allowing them to relax.
For the system in Fig.~\ref{fig:4},
if we consider a much lower
oscillation frequency or allow the system to remain
at $B = 0.0$ for an extended period of time,
the gradient in the stripe patterns that appear
Fig.~\ref{fig:4}(d) begins to flatten out over time.
When the oscillation frequency is finite, however,
as $B$ begins to increase from zero the pattern contracts back into
a bubble state, and there
is not enough time for the gradient to flatten.
The fact that the asymmetric pattern can relax causes the ratchet effect
to be strongly frequency dependent, as we discuss later.

\begin{figure}
\includegraphics[width=\columnwidth]{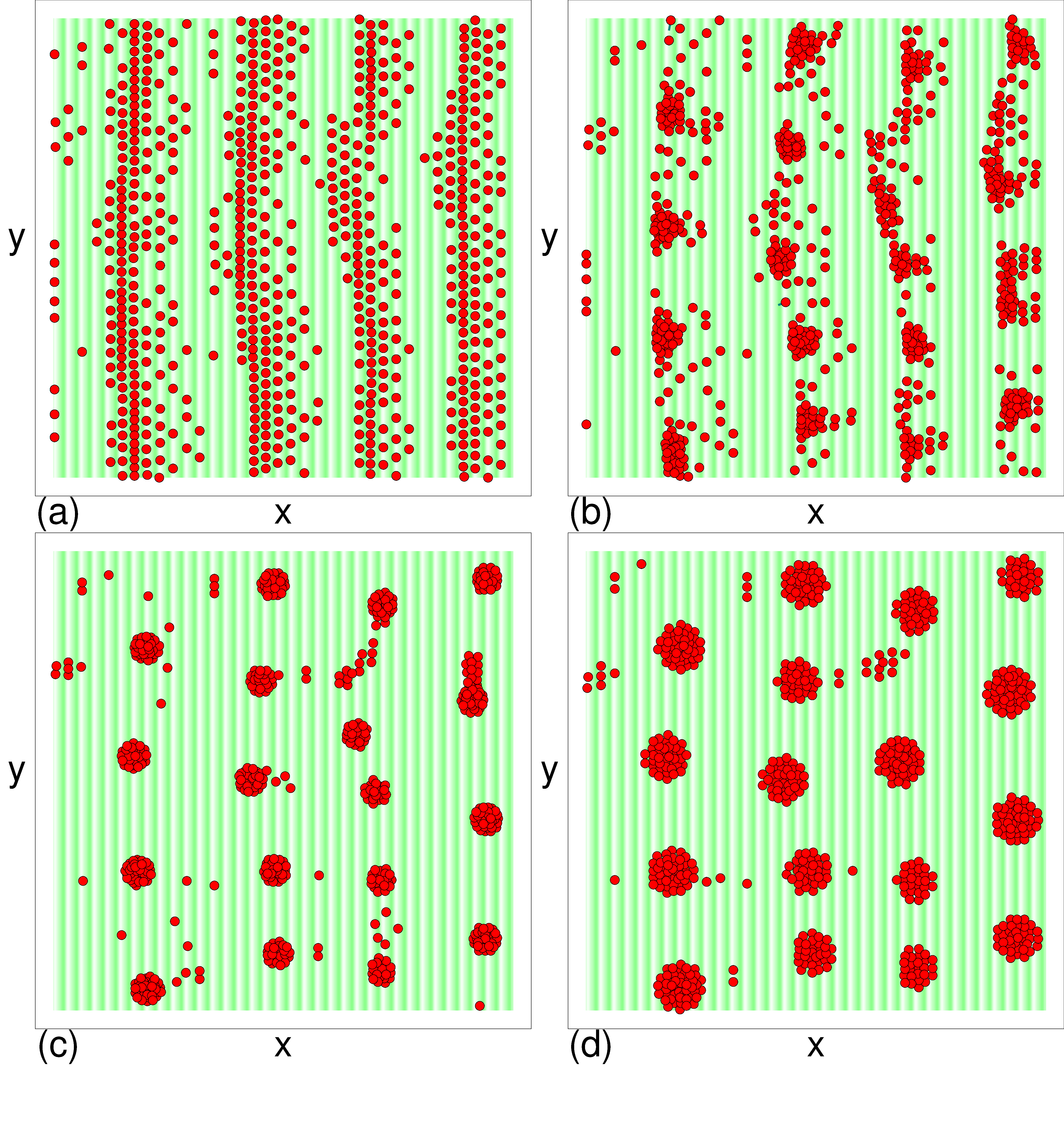}
\caption{Images of particle positions (red) and the substrate
potential (green) taken during the second half of a cycle 
with $B$ increasing from $B=0.0$ to $B = 4.2$ as well as during the next cycle
for the system in Fig.~\ref{fig:3} with $A_p=2.0$, $N_p=35$,
$\omega/2\pi=2.5 \times 10^{-5}$, and $\rho=0.52$.
(a) $B = 1.0$, showing the stripe pattern beginning to contract.
(b) At $B = 3.2$, the stripes are collapsing into bubbles.
(c) The bubble state just before $B$ reaches its maximum value of $B = 4.2$.
(d) The next expansion cycle at $B = 2.8$, after the first expansion and
contraction cycle has completed.  
}
\label{fig:5}
\end{figure}

In Fig.~\ref{fig:5}, we show the continuation
of the second half of the cycle from Fig.~\ref{fig:4},
where $B$ is increasing from $B=0.0$ back up to $B=4.2$.
At $B=1.0$ in Fig.~\ref{fig:5}(a), the pattern has begun to contract,
while at $B=3.2$ in Fig.~\ref{fig:5}(b), the stripes begin to collapse
into bubbles.
The bubbles then become more compact, as shown in
Fig.~\ref{fig:5}(c) at a point just before
$B$ reaches its maximum value of $B = 4.2$.
Figure~\ref{fig:5}(d) shows the $B=2.8$ state during the
expansion phase of the next cycle as $B$ is reduced again; the
configuration is very similar to that shown in Fig.~\ref{fig:4}(b).

Figures~\ref{fig:4} and \ref{fig:5} indicate that
the expansion and collapse of the bubbles occurs in an asymmetric fashion
due to the presence of the substrate.
By the time $B$ reaches $B = 0.0$, as shown in Fig.~\ref{fig:4}(d),
the particles are asymmetrically distributed since
particles that are moving in the negative $x$ or hard direction travel
more slowly, and thus translate by a shorter distance, than
particles that are moving in the positive $x$ or
easy direction.
Before the particles moving in the negative direction are able to
relax and diminish the density gradient that has formed,
$B$ begins to increase again, and as a result,
when the bubbles reform, their positions have been
shifted in the positive $x$ direction.

\begin{figure}
\includegraphics[width=\columnwidth]{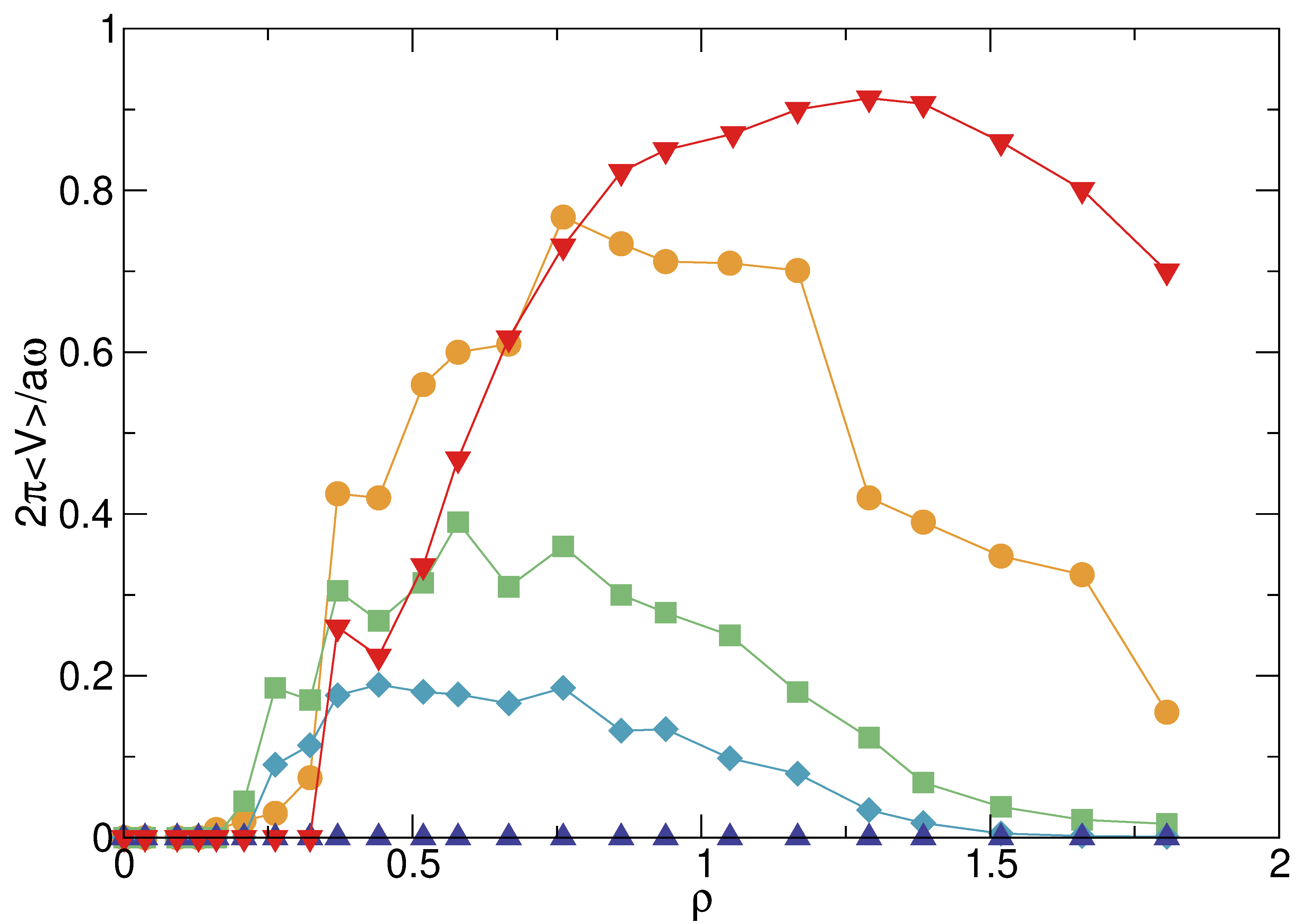}
\caption{The normalized ratchet velocity,
$2\pi\langle V\rangle/a\omega$ vs particle density $\rho$ for
systems with $N_p=35$ and $\omega/2\pi=2.5\times 10^{-5}$ at
$A_p = 0.0$ (dark blue up triangles),
$0.75$ (light blue diamonds),
$1.0$ (green squares),
$2.0$ (orange circles),
and $3.2$ (red down triangles).
A critical density of particles is required
for the ratchet effect to occur, and
there is an optimal density for ratcheting that varies with $A_p$.
}
\label{fig:6}
\end{figure}

Figure~\ref{fig:6}
shows the normalized
ratchet velocity $2\pi\langle V\rangle/a\omega$
versus particle density $\rho$ for
samples with $N_p=35$ and $\omega/2\pi=2.5\times 10^{-5}$ at
$A_p = 0.0$, 0.75, 1.0, 2.0, and 3.2.
For this value of $N_p$, the substrate spacing is
$a=1.01$,
and with the normalization, the ratchet
is measured in terms of the average number of
substrate minima by which the particles translate during each
cycle.
The maximum ratchet efficiency is close to one
substrate spacing per cycle in Fig.~\ref{fig:6},
but in general, the ratchet efficiencies we observe are considerably
less than one substrate spacing per cycle.
The ratchet efficiency
decreases as $A_p$ is reduced,
and no ratchet effect occurs when $A_p = 0.0$.
When $A_p$ is finite,
there is a critical density for the ratchet effect to occur that
falls between $\rho=0.2$ and $\rho=0.8$, and
there is also an optimal density at which the ratchet efficiency is
maximized.
At low particle densities, bubbles are unable to form since the distance
between adjacent particles is too large, and the ratchet effect is
lost.
At higher densities, the bubbles become large, causing the particles
to spread out more rapidly
during the $B = 0.0$ portion of the cycle.
This diminishes the density gradients that are present, and
reduces the ratchet effect.
In many ratchet systems, there is a quantization of the motion when
the particles
move by an integer number of lattice spacings per our cycle.
In our system, however, the motion is much more disordered,
as shown by the trajectories in
Fig.~\ref{fig:3}, washing out any quantization effects.

\begin{figure}
\includegraphics[width=\columnwidth]{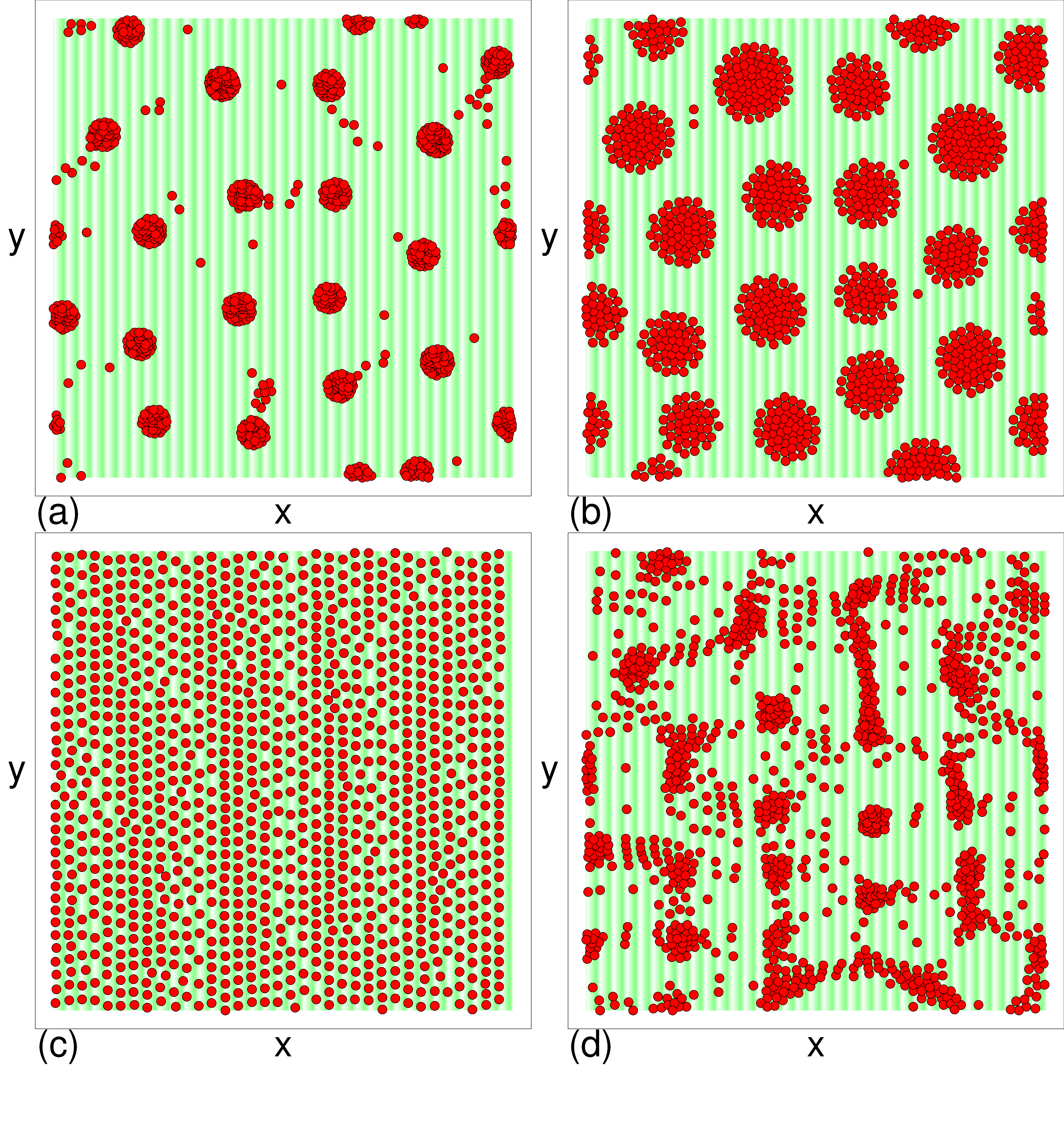}
\caption{Images of particle positions (red) and the substrate
potential (green)  
during a single cycle for
a system with $N_p=35$, $\omega/2\pi=2.5\times 10^{-5}$,
$A_p = 0.75$ and $\rho = 0.86$,
where the ratchet efficiency is reduced.
(a) Compact bubbles at $B = 4.2$.
(b) Reduction of $B$ to $B = 2.4$, showing large bubbles that are expanding.
(c) At $B = 0$, a weak density gradient is present.
(d) After $B$ has increased back to $B = 2.8$,
the bubbles begin to reform.}
\label{fig:7}
\end{figure}

For smaller $A_p$, we find that the density gradients
in the $B = 0.0$ portion of the cycle are diminished.
In Fig.~\ref{fig:7}, we show images of the system
from Fig.~\ref{fig:6} with $A_p = 0.75$ and $\rho = 0.86$, which has a
reduced ratchet effect compared to the
$A_p = 2.0$ and $\rho = 0.52$ system from Fig.~\ref{fig:4}.
When $B=4.2$ at the beginning of the cycle,
small bubbles form as shown in Fig.~\ref{fig:7}(a).
After $B$ has decreased to $B=2.4$,
Fig.~\ref{fig:7}(b) indicates that the bubbles have become much larger,
and are expanding considerably further than in the $A_p=2.0$ system.
At the $B=0.0$ state in Fig.~\ref{fig:7}(c), a nearly uniform crystal has
formed with only weak particle density gradient stripes present.
When $B$ increases again,
Fig.~\ref{fig:7}(d) shows that at $B = 2.8$,
the bubbles begin to reform.
This result shows that it is the formation of a gradient at $B=0.0$ that
gives rise to the positive ratchet effect, and when this gradient becomes
weaker, the ratchet effect also becomes weaker.

\begin{figure}
\includegraphics[width=\columnwidth]{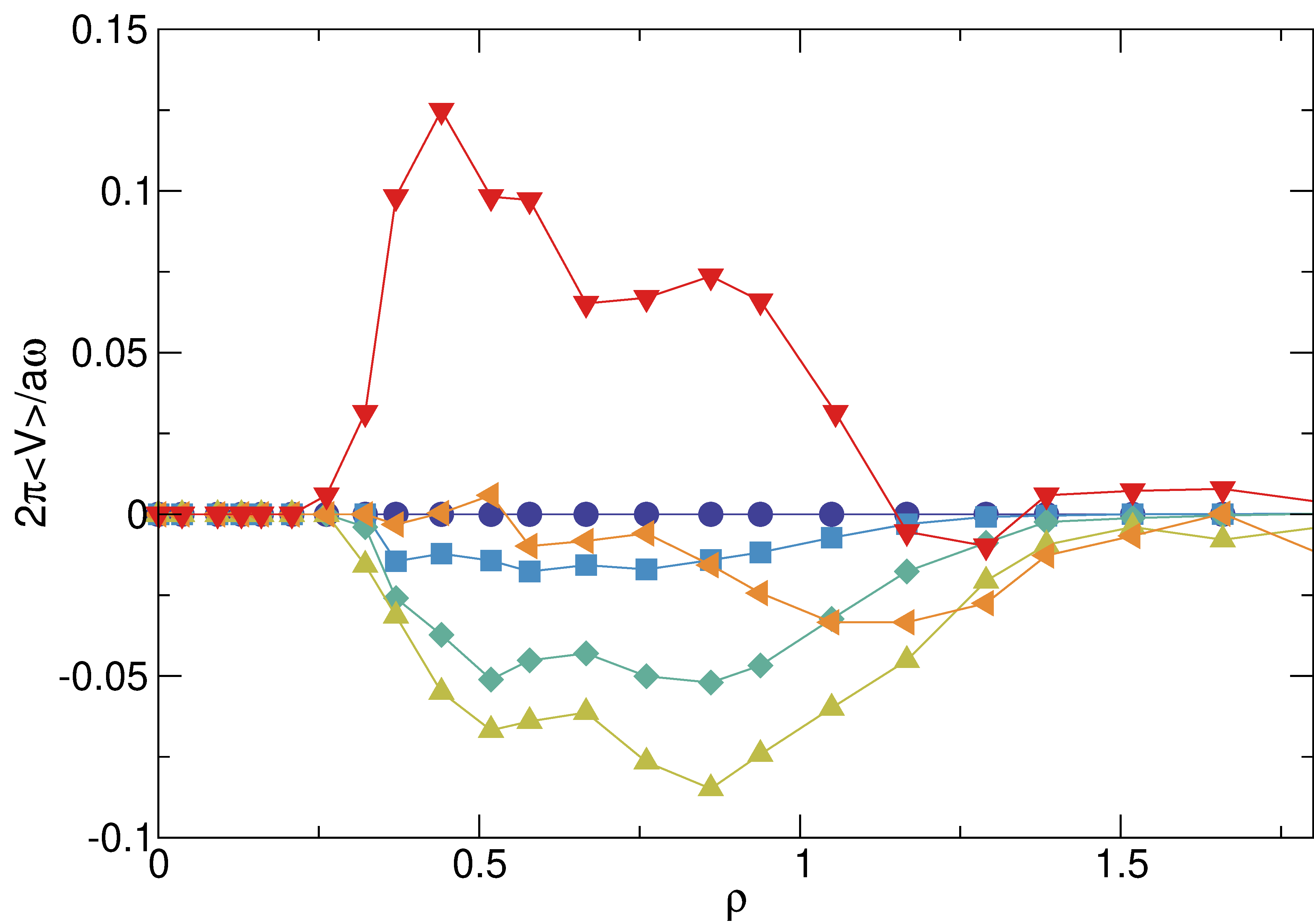}
\caption{The averaged ratchet velocity per particle $2\pi\langle V\rangle/a\omega$ vs particle density $\rho$ for systems with $N_p=35$ and
  $\omega/2\pi=2.5 \times 10^{-5}$ at
$A_p = 0.5$ (red down triangles), 0.375 (orange left triangles),
0.25 (green up triangles), 0.125 (teal diamonds), 0.0625 (light blue squares),
and 0.0 (dark blue circles).
There is a reversal of the ratchet effect for $A_p < 0.5$.
}
\label{fig:8}
\end{figure}

For $A_p < 0.5$, we find a reversed ratchet effect, as shown in
Fig.~\ref{fig:8} where we plot $2\pi\langle V\rangle/a\omega$ versus
$\rho$ for samples with $N_p=35$ and $\omega/2\pi=2.5\times 10^{-5}$ at
$A_p = 0.5$, 0.375, 0.25, 0.125, 0.0625, and 0.0.
At $A_p = 0.5$, the ratchet effect is still positive,
but the ratchet motion becomes negative
for $A_p = 0.25$. The magnitude of the ratchet effect then
diminishes with decreasing $A_p$ until it reaches zero for $A_p=0.0$.
In general, the ratchet efficiency is much lower for
the reversed ratchet motion than for the forward ratchet
motion for these values of $N_p$ and $\omega$.

\begin{figure}
\includegraphics[width=\columnwidth]{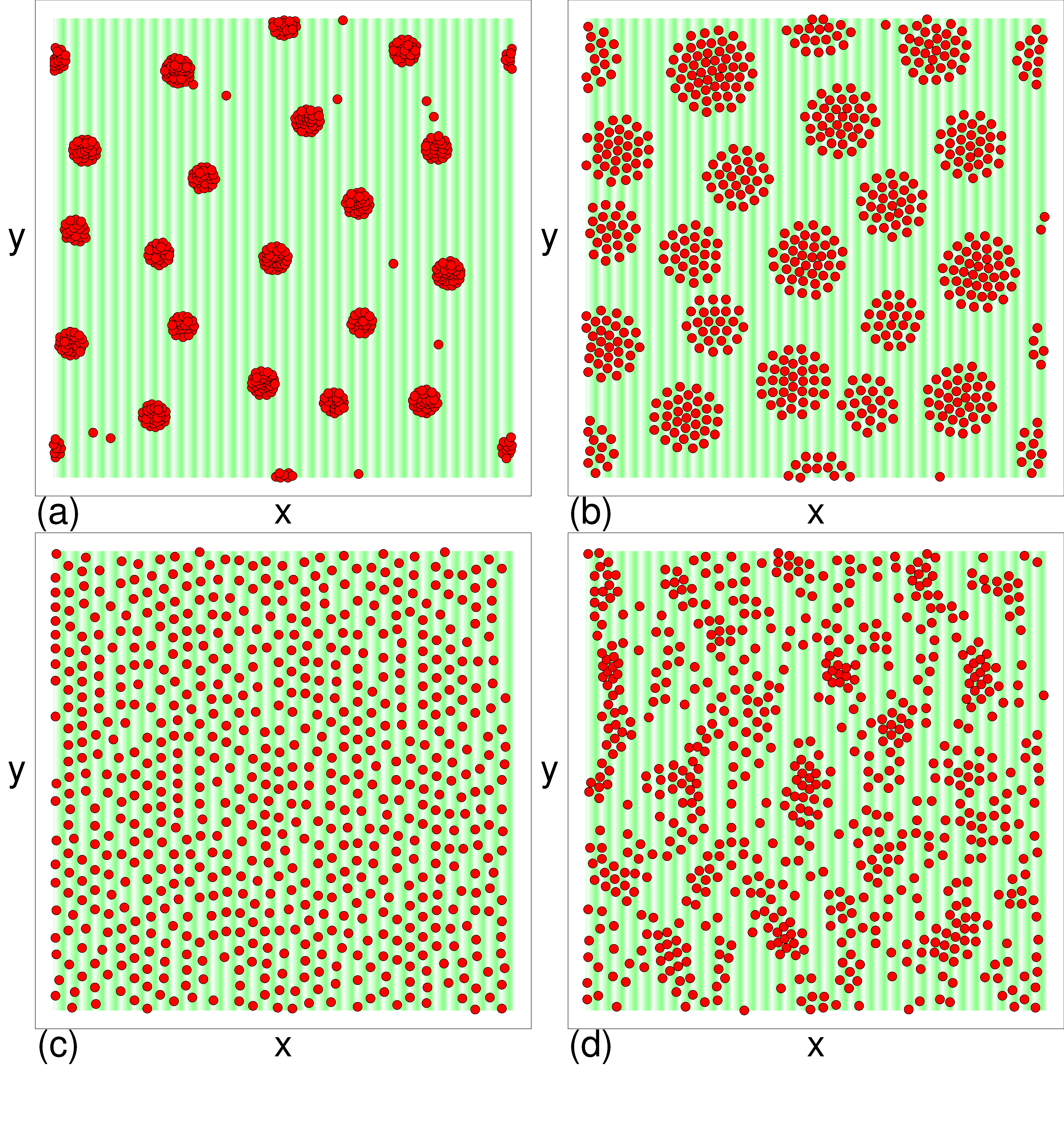}
\caption{Images of particle positions (red) and the substrate
potential (green) during a single cycle
for a system with $N_p=35$, $\omega/2\pi=2.5 \times 10^{-5}$,
$A_p = 0.25$, and $\rho = 0.52$, where the ratchet effect is negative.
(a) At $B = 4.2$, the bubbles are compact.
(b) When $B$ decreases to $B = 2.0$,
larger bubbles appear that are expanding rapidly.
(c) At $B = 0.0$, the density is mostly uniform with no
significant gradient.
(d) When $B$ increases back to $B = 2.8$, the bubbles start to reform.
}
\label{fig:9}
\end{figure}

In Fig.~\ref{fig:9}, we show the particle positions at different points
in the cycle for a sample with $N_p=35$, $\omega/2\pi=2.5\times 10^{-5}$,
$A_p=0.25$, and $\rho=0.52$, where there is a reversed ratchet effect.
At $B=4.2$ in Fig.~\ref{fig:9}(a), the bubbles are compact and are
interspersed with a small number of particles that did not get trapped
in a bubble.
In Fig.~\ref{fig:9}(b), when $B$ decreases to $B = 2.0$,
large, rapidly expanding bubbles appear, and the asymmetry
observed in the bubbles at higher $A_p$ is lost.
At $B=0.0$ in Fig.~\ref{fig:9}(c), the particles are spread fairly uniformly
across the system with no significant density gradient.
During the contraction phase of increasing $B$,
Fig.~\ref{fig:9}(d) shows that at $B = 2.8$,
the bubbles are beginning to reform.
Since there is no density gradient at $B=0.0$, the reversed ratchet
effect arises as a result of
the motion of the individual particles that got left
out of the reformed bubbles.
During the increasing $B$ portion of the cycle,
the bubbles rapidly collapse with the center of each bubble located
at a substrate potential minimum. As a result of the substrate asymmetry,
particles moving in the positive $x$-direction toward the bubble travel
slightly slower than particles moving in the negative $x$-direction
toward the bubble.
If the bubble collapse is rapid enough, some of the slower-moving particles
cannot keep up with the collapse and end up becoming
trapped by the substrate without joining a bubble,
leading to the negative ratchet effect.

\begin{figure}
\includegraphics[width=\columnwidth]{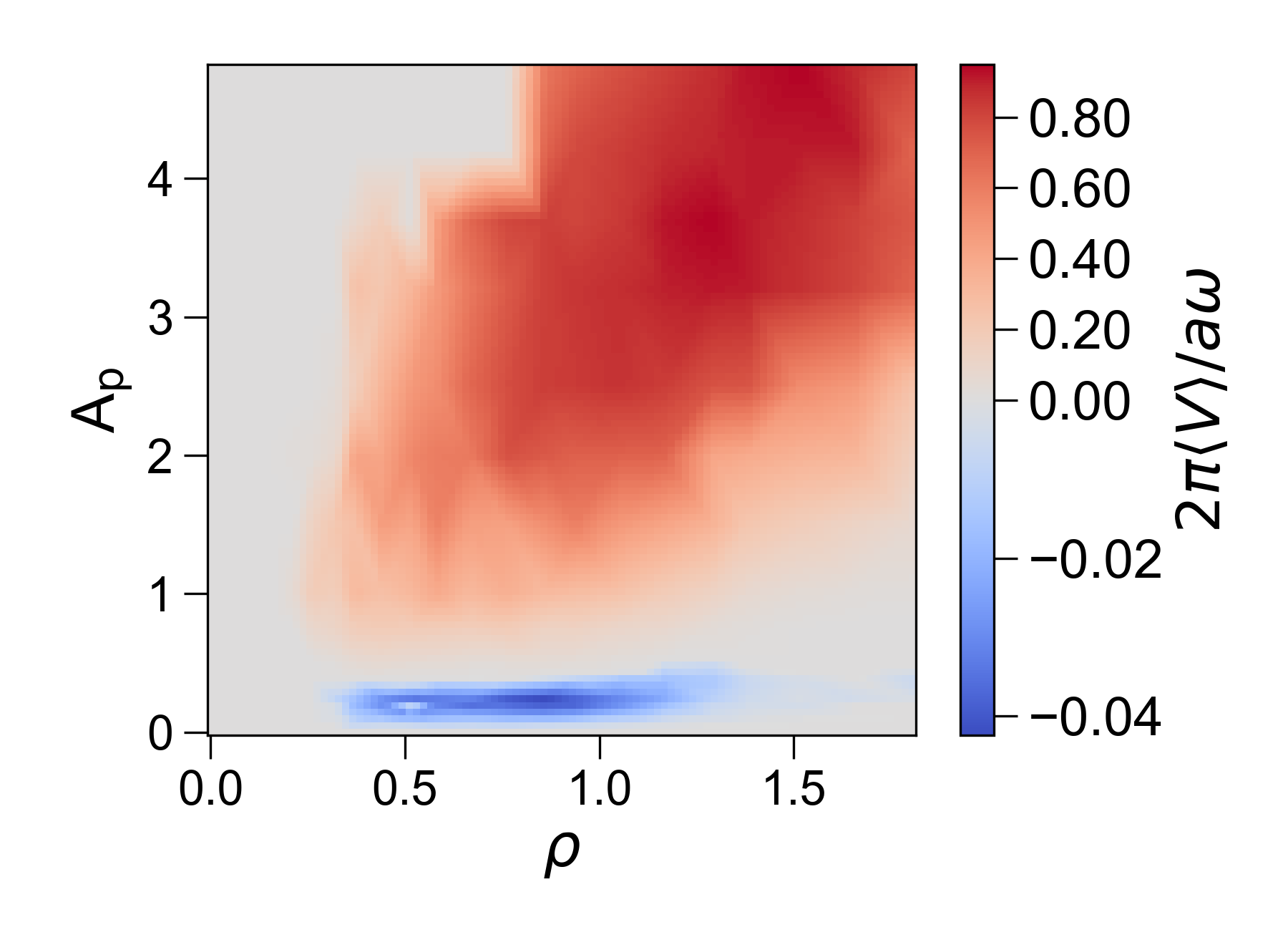}
\caption{Heatmap of $2\pi\langle V\rangle/a\omega$ as a function of
$A_p$ vs $\rho$ in a system with $N_p=35$ and $\omega/2\pi=2.5\times 10^{-5}$,
highlighting the regimes of positive (red) and negative
(blue) ratchet effects.
}
\label{fig:10}
\end{figure}

\begin{figure}
\includegraphics[width=\columnwidth]{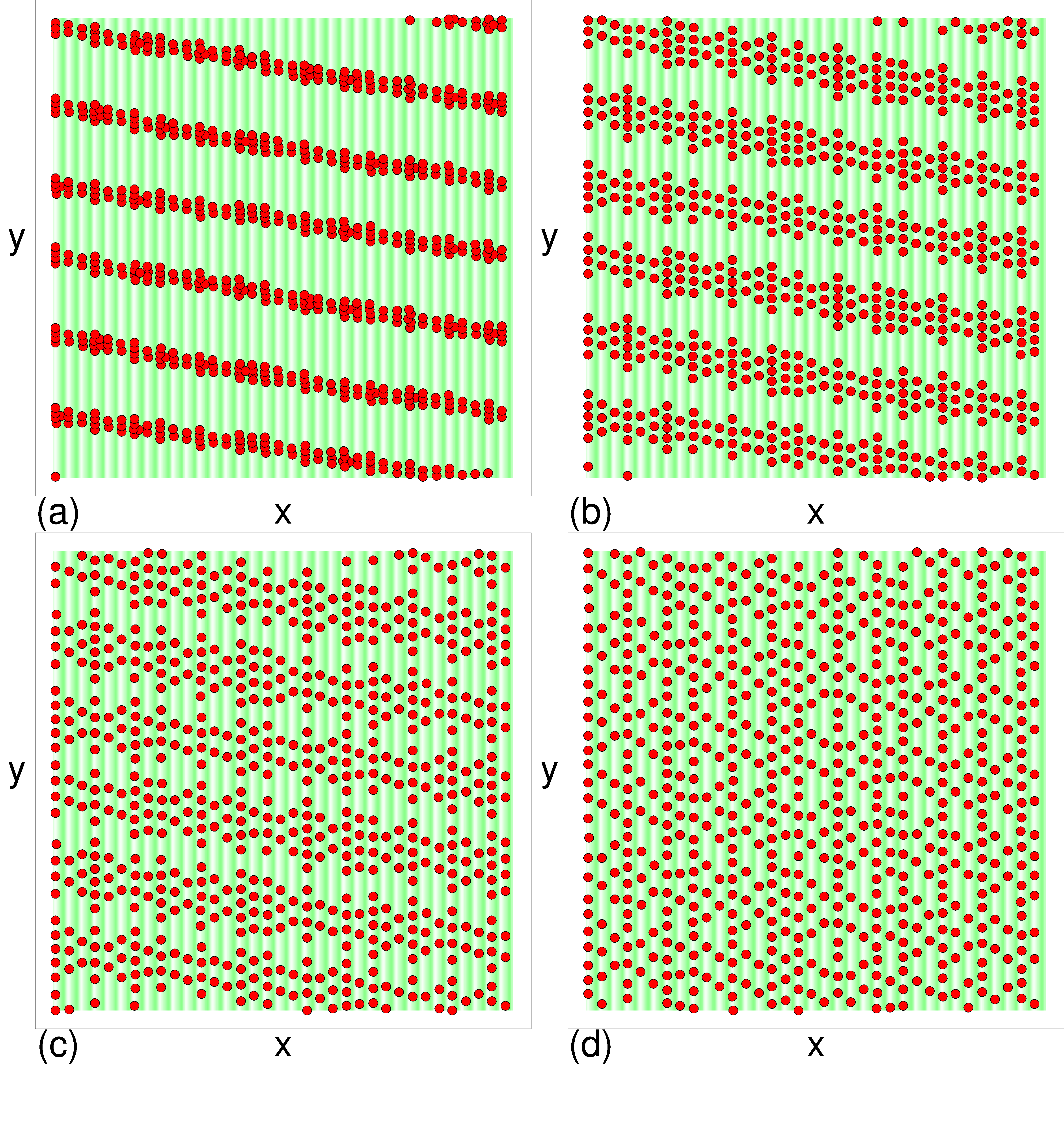}
\caption{Images of particle positions (red) and the substrate
potential (green) during the decreasing $B$ portion of a cycle for
a system with $N_p=35$, $\omega/2\pi=2.5\times 10^{-5}$,
$A_{p} = 4.0$, and $\rho = 0.52$,
where there is no ratchet effect and the particles can move only in the
$y$ direction.
(a) Pinned stripes at $B = 4.2$.
(b) $B = 1.8$, where the stripes are expanding. 
(c) Uniform stripes at $B = 0.6$.
(d) Uniform modulated phase at $B = 0.0$.
	}
\label{fig:11}
\end{figure}

In Fig.~\ref{fig:10}, we show a heat diagram
of $\langle V\rangle$ as a function of $A_p$ versus $\rho$ for
a system with $N_p=35$ and $\omega/2\pi=2.5\times 10^{-5}$,
highlighting the regimes of positive and negative ratchet effects.
When $\rho < 0.2$, no ratchet effect appears,
while there is a strong
positive ratchet effect for large $A_p$ and large $\rho$.
For a fixed $\rho = 0.52$,
as $A_p$ increases there is initially a negative ratchet effect that
transitions to a positive ratchet effect,
while at even higher $A_p$ the ratchet effect is lost once
the particles become strongly pinned in the $x$-direction and are no longer
able to jump between substrate minima
as $B$ is cycled.
In Fig.~\ref{fig:11},
we illustrate the particle positions in the decreasing $B$ portion of the
cycle
for the system from Fig.~\ref{fig:10} at $\rho = 0.52$ and $A_p = 4.0$,
where there is no ratchet effect.
At $B=4.2$ in Fig.~\ref{fig:11}(a), a pinned stripe structure forms.
As $B$ decreases, the particles can only move
in the positive or negative $y$-direction,
and the stripe expands, as shown in Fig.~\ref{fig:11}(b) at $B = 1.8$.
Eventually the particles become more uniformly distributed across
the sample, as indicated in Fig.~\ref{fig:11}(c) at $B = 0.6$
and in Fig.~\ref{fig:11}(d) at $B = 0.0$.
For lower $\rho$ but stronger $A_p$, a ratchet effect does not occur,
and in general, a ratchet effect only arises when the particles
are able to form bubbles.

\begin{figure}
\includegraphics[width=\columnwidth]{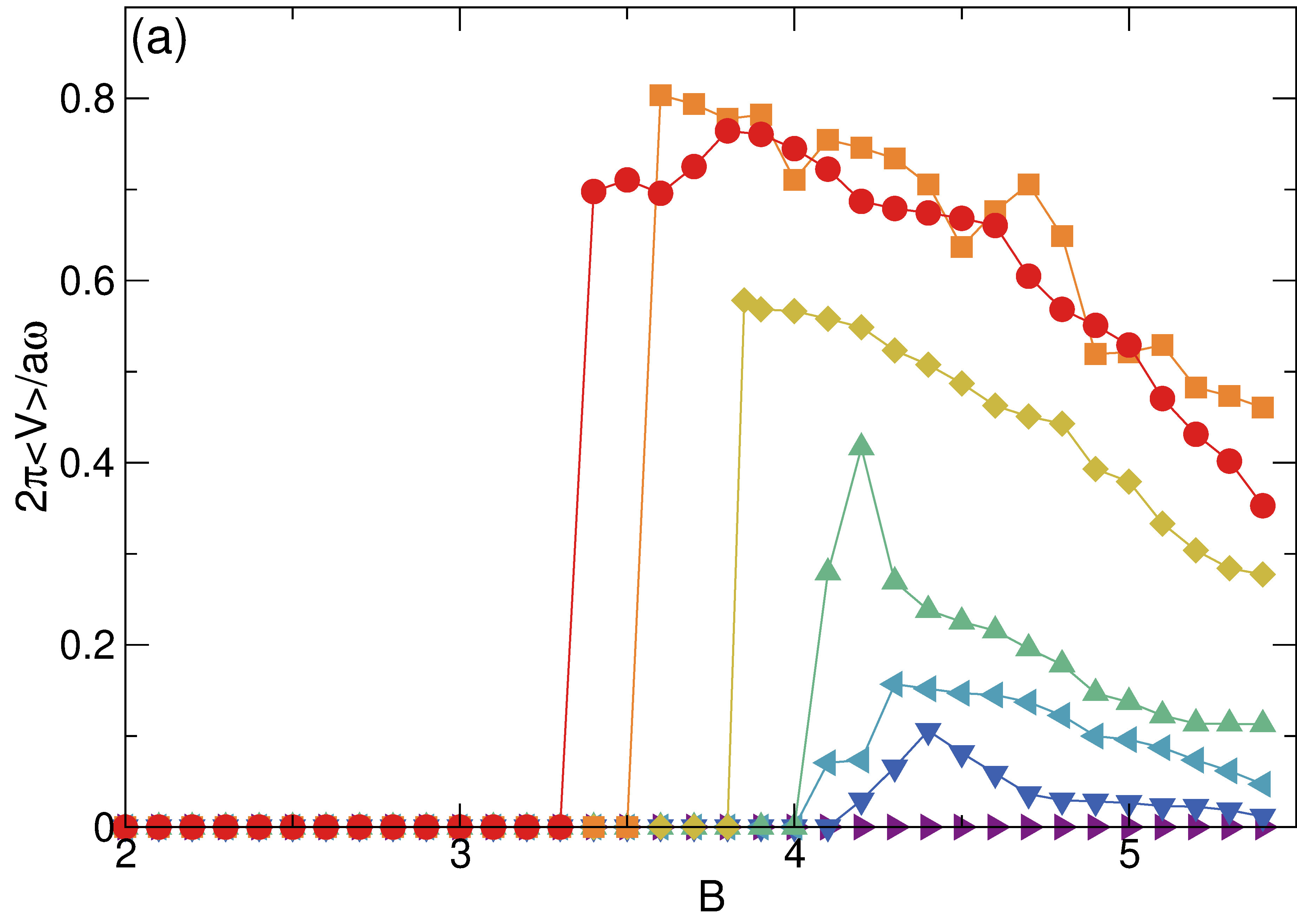}
\includegraphics[width=\columnwidth]{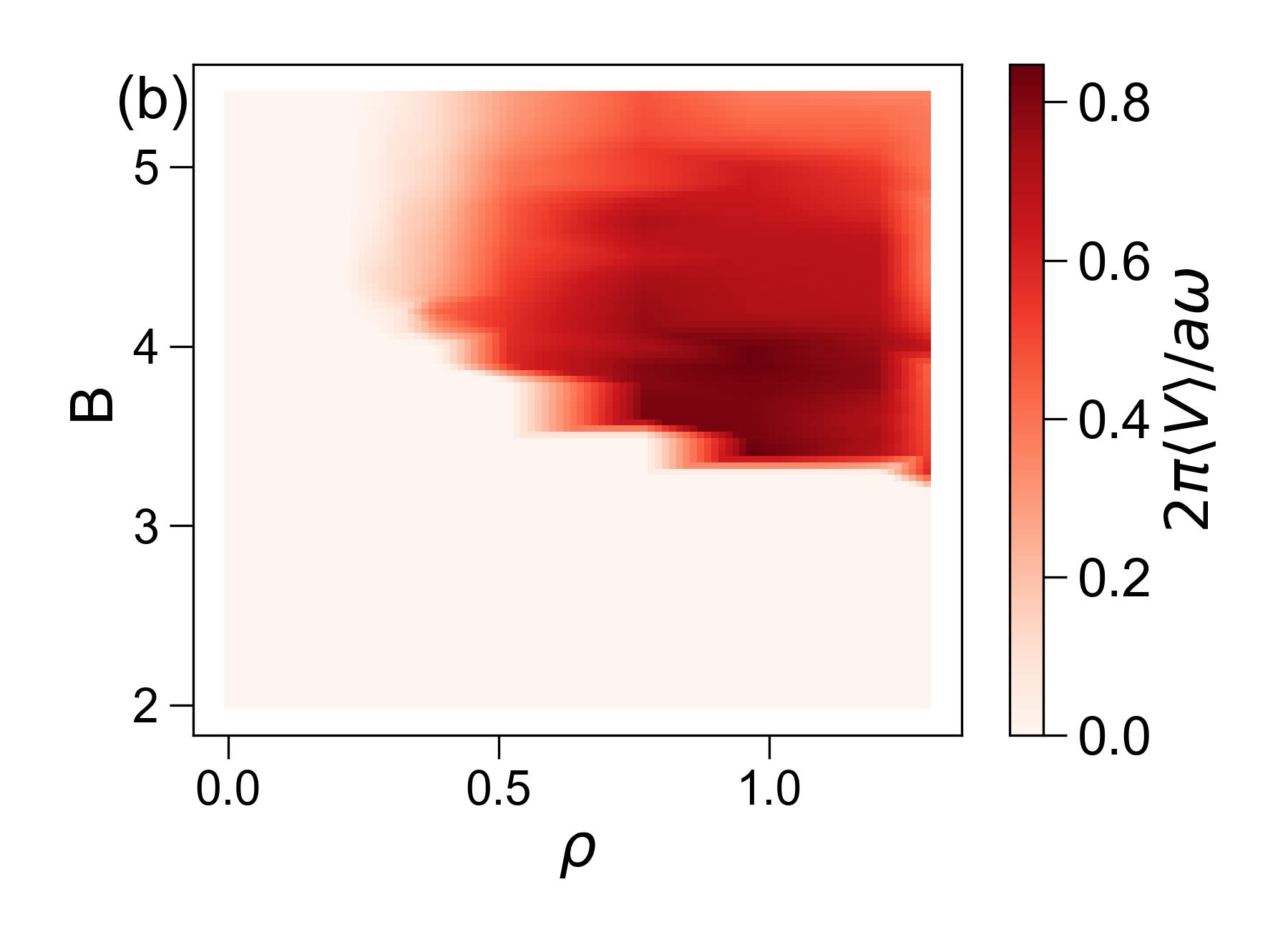}
\caption{(a) Average ratchet velocity $2\pi\langle V\rangle/a\omega$
vs $B$ for samples with $N_p=35$ and $\omega/2\pi=2.5\times 10^{-5}$ at
$\rho = 1.2$ (red circles),
$0.8$ (orange squares),
$0.52$ (pale green diamonds),
$0.4$ (teal up triangles),
$0.35$ (light blue left triangles),
$0.3$ (dark blue down triangles),
and $0.2$ (purple right triangles).
There is a critical $B$ value
for the ratchet effect to occur.
(b) Heatmap of $2\pi\langle V\rangle/a\omega$ as a function of
$\rho$ vs $B$ for the same system
showing where the ratchet effect occurs.
}
\label{fig:12}
\end{figure}

We next consider the effect of changing the maximum $B$
for a system with fixed $A_p = 2.0$.
In Fig.~\ref{fig:12}(a), we plot the
ratchet velocity $2\pi\langle V\rangle/a\omega$ versus $B$
in a sample with $N_p=35$ and $\omega/2\pi=2.5\times 10^{-5}$ at
$\rho = 1.2$, $0.8$, $0.52$, $0.4$, $0.35$, $0.3$, and $0.2$.
We find that there is a critical $B$ value for the ratchet
effect to occur, and the maximum
ratchet efficiency occurs just above this critical field.
For these parameters, we do not observe a reverse ratchet effect.
In Fig.~\ref{fig:12}(b), we show a
heatmap of the ratchet velocity
as a function of $\rho$ versus $B$, indicating where the ratchet effect occurs.

\section{Frequency Dependence}

\begin{figure}
\includegraphics[width=\columnwidth]{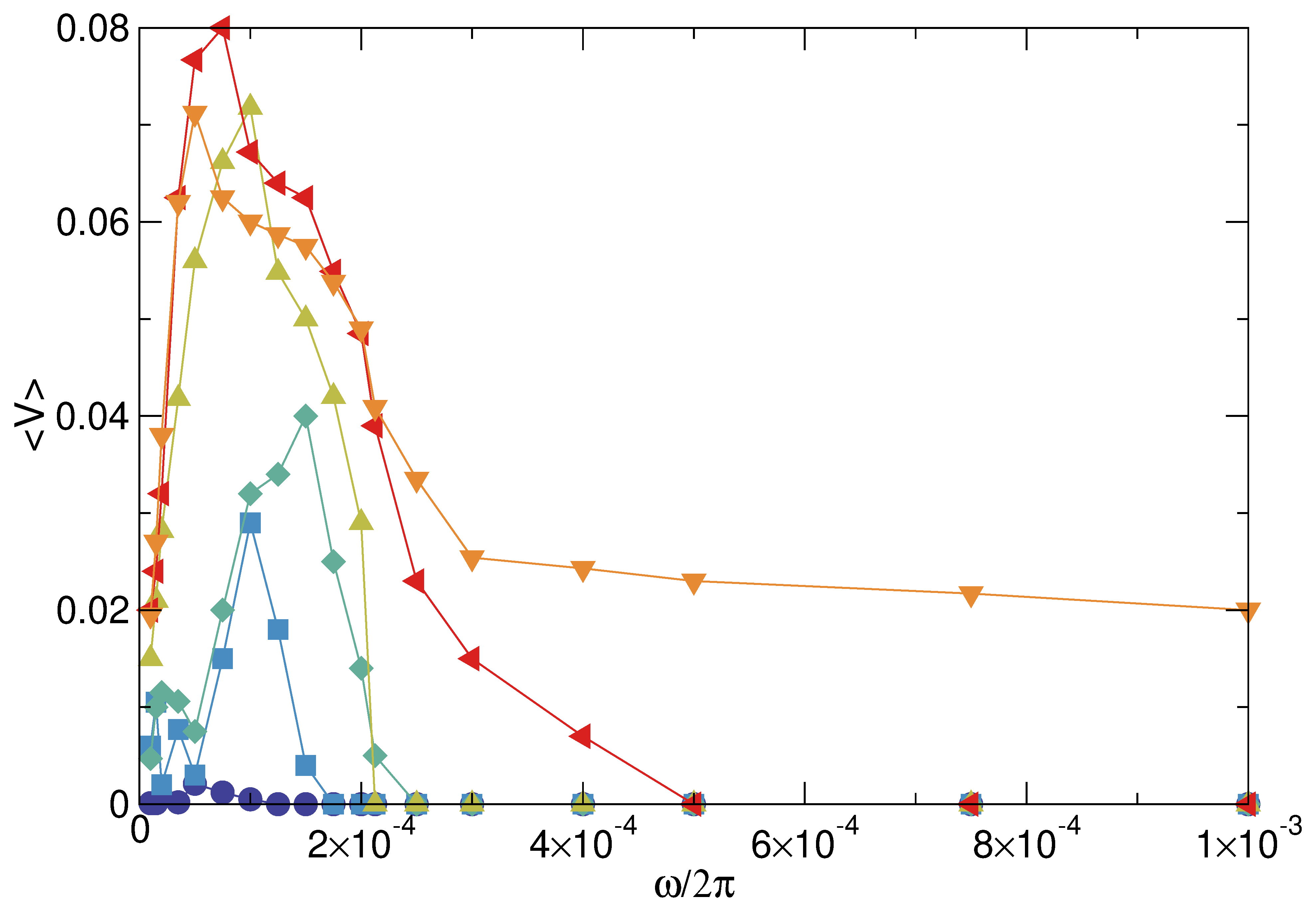}
\caption{The ratchet velocity $\langle V\rangle$ vs $\omega/2\pi$ for
a sample with $N_p=35$ and $A_p=2.0$ at
$\rho=0.16$ (dark blue circles),
$0.26$ (light blue squares),
$0.32$ (teal diamonds),
$0.52$ (pale green up triangles),
$0.76$ (orange down triangles),
and $0.938$ (red left triangles).
	}
\label{fig:13}
\end{figure}

\begin{figure}
\includegraphics[width=\columnwidth]{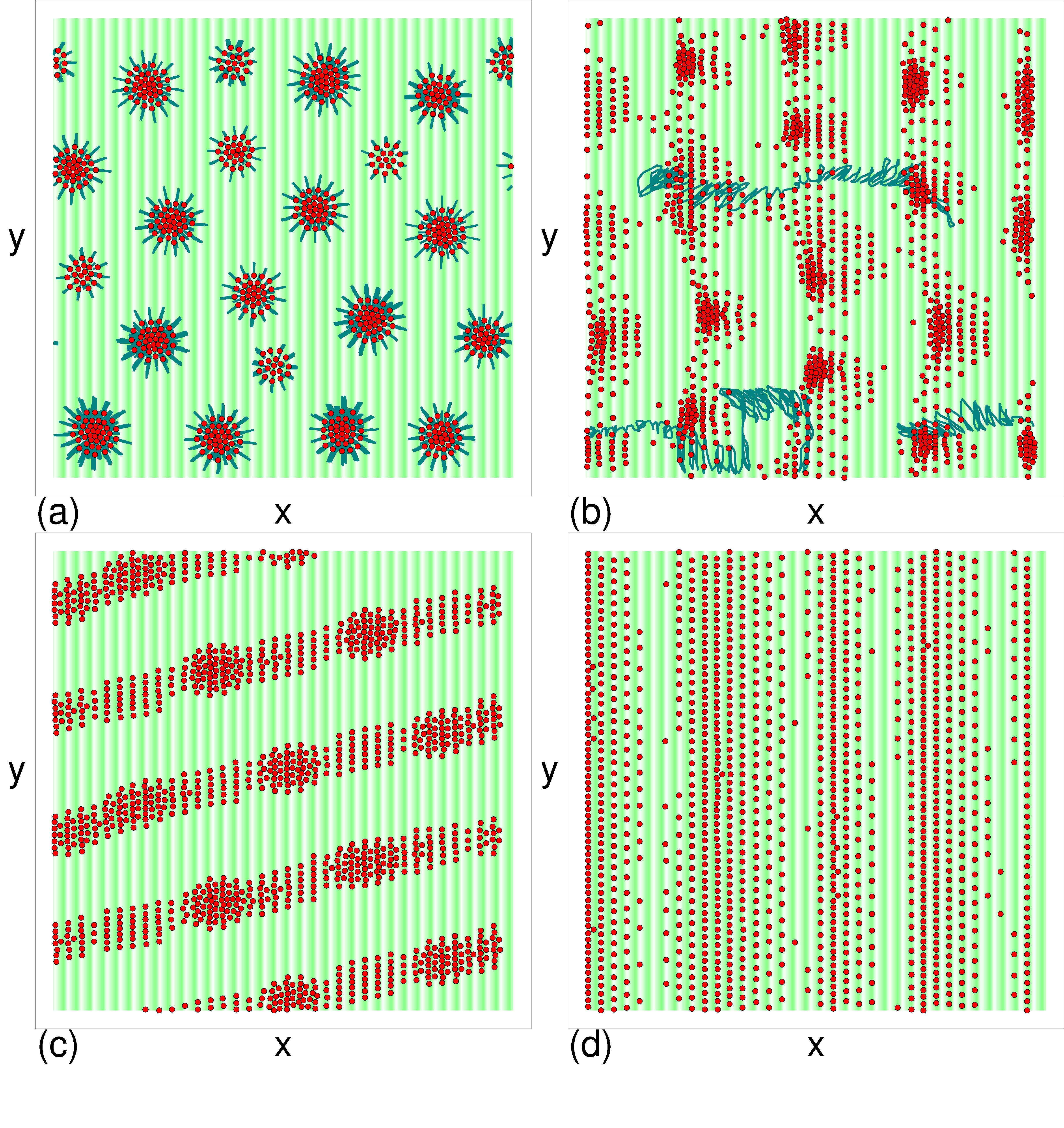}
\caption{Image of particle positions (red) and the underlying
asymmetric potential (green) in samples with $N_p=35$.
(a) Trajectories of all particles (lines) during 20 oscillation
cycles at $\rho = 0.52$, $\omega/2\pi = 2.5 \times 10^{-4}$,
and $A_p = 0.0$,
showing that at high frequencies the system remains
in an oscillating bubble state.
(b) Trajectories of two particles (lines) during 20 oscillation
cycles at $\rho = 0.938$, $\omega/2\pi = 7.5 \times 10^{-5}$,
and $A_p = 3.7$ for the system shown in Fig.~\ref{fig:15}(a)
where there is a strong ratchet effect.
(c) Particle positions for the sample from
panel (b) at $\omega/2\pi = 1.25\times 10^{-3}$,
where there is no ratchet effect and the system forms a
pinned modulated stripe state.
(d) Particle positions for the sample from panel (b)
at $\omega/2\pi = 5 \times 10^{-6}$ during
the $B = 0.0$ portion of the cycle,
revealing a more uniform stripe-like pattern.
}
\label{fig:14}
\end{figure}

We next vary the frequency of the oscillations. In Fig. 13, we plot $\langle V \rangle$ versus $\omega/2\pi$ for
a system with $N_p=35$ and $A_p=2.0$ at
$\rho = 0.16$, $0.26$, $0.32$, $0.52$, $0.76$, and $0.938$.
At $\rho = 0.16$, there is almost no ratchet effect,
but the maximum ratchet efficiency increases with increasing $\rho$.
There is an optimal frequency in each case; for example,
at $\rho = 0.52$, the optimum frequency falls at
$\omega/2\pi=1 \times 10^{-4}.$
For higher frequencies, the oscillation becomes fast enough
that the particles can no longer respond to it,
and the system remains in an oscillating bubble pattern,
as shown in Fig.~\ref{fig:14}(a)
for a sample with $\rho = 0.52$, $\omega/2\pi = 0.00025$,
and $A_p = 0.0$.

In Fig.~\ref{fig:13}, for $\rho<0.76$
the ratchet effect is lost when
$\omega > 2 \times 10^{-4}$,
while as the particle density approaches $\rho = 1.0$,
the ratchet effect can persist up to much higher frequencies
because the bubbles become
large enough to percolate across the sample
during the $B = 0.0$ portion of the cycle.
For very small frequencies, 
the ratchet efficiency is reduced
since the particles have more time to
relax into a more uniform state during the $B = 0.0$ portion of the cycle.
We have also examined the
effect of changing the frequency
for other values of $\rho$ and $A_p$,
and find results similar to those shown in Fig.~\ref{fig:13}.

\begin{figure}
\includegraphics[width=\columnwidth]{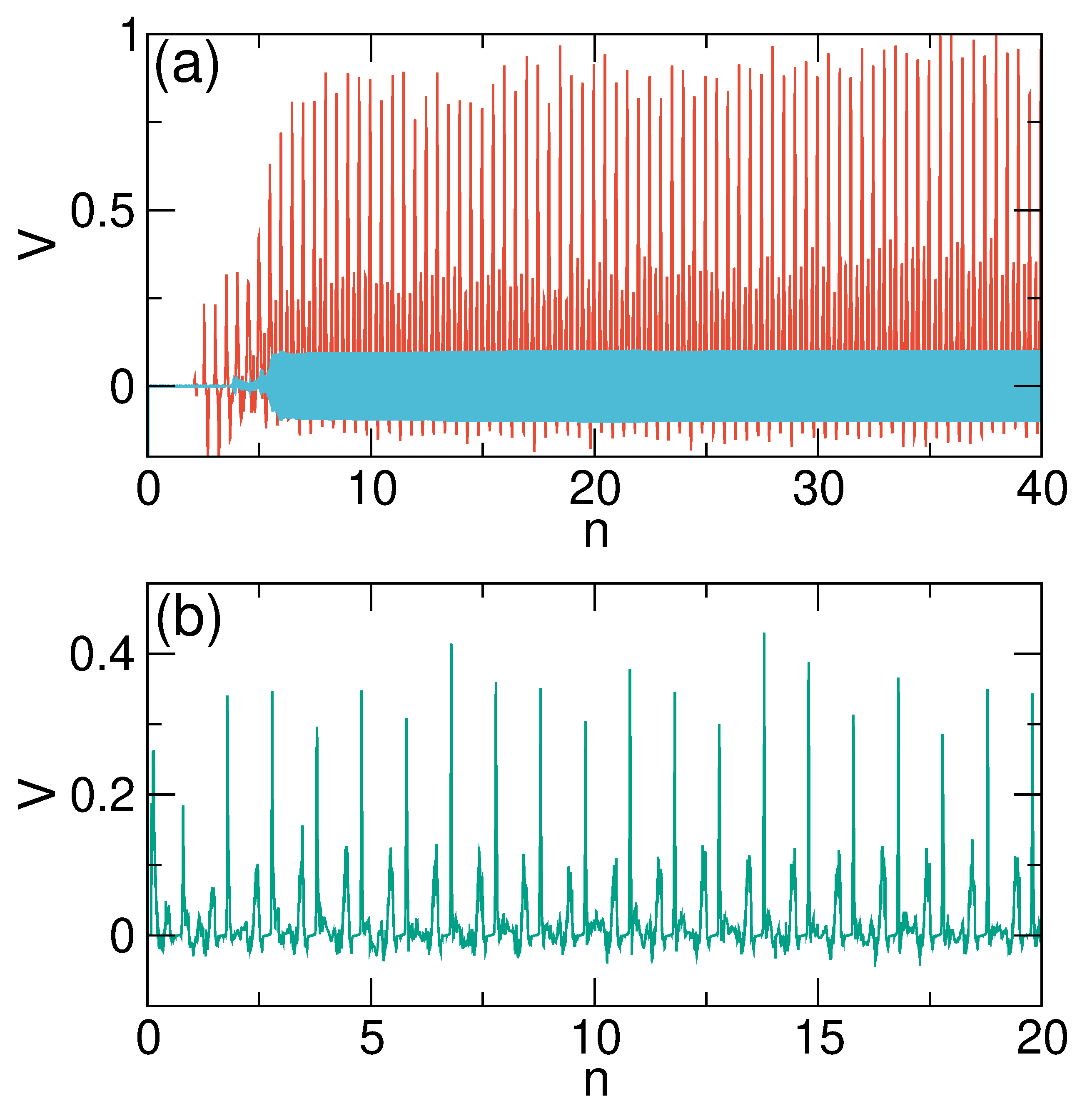}
\caption{The instantaneous average velocity $V$ vs time in cycle
numbers $n$
for a system with  
$\rho = 0.938$, $N_p=35$, and $A_{p} = 3.7$ at
$\omega/2\pi = 7.5\times 10^{-5}$ (red) and
$\omega/2\pi = 1.25 \times 10^{-3}$ (blue).
The period on the $x$-axis is measured in terms of
the $\omega/2\pi = 7.5\times 10^{-5}$ system.
There is a strong ratchet effect for $\omega/2\pi = 7.5\times 10^{-5}$ and no
ratchet effect for
$\omega/2\pi = 1.25\times 10^{-3}$.
(b) The same at $\omega/2\pi = 5 \times 10^{-6}$.}
\label{fig:15}
\end{figure}

In Fig.~\ref{fig:15}(a), we plot the time series
of the instantaneous velocity $V$
for a system with $\rho=0.938$, $N_p=35$, and $A_p = 3.7$
at $\omega/2\pi = 7.5\times 10^{-5}$ and
$\omega/2\pi = 1.25\times 10^{-3}$.
When $\omega/2\pi=7.5\times 10^{-5}$,
there is a strong positive velocity response
with maximum velocity values of nearly $V=0.9$,
while for the $\omega/2\pi = 1.25\times 10^{-3}$
system, there is no ratchet effect,
and the velocity oscillates around zero.
Figure~\ref{fig:14}(b)
shows the particle positions and the trajectories of
two representative particles
for the system from
Fig.~\ref{fig:15}(a) at $\omega/2\pi = 7.5\times 10^{-5}$,
where there is a ratchet effect and the velocities are predominantly
in the positive $x$ direction.
In Fig.~\ref{fig:14}(c), we illustrate
the particle configurations for
the system from Fig.~\ref{fig:15}(a)
with $\omega/2\pi = 1.25\times 10^{-3}$,
where we find a pinned stripe state with some density modulations.
Figure~\ref{fig:15}(b) shows the
instantaneous velocity time series for
the same system from Fig.~\ref{fig:15}(a)
but for a frequency that is 15 times smaller, $\omega/2\pi = 5\times 10^{-6}$.
The maximum velocity peaks at around $V=0.35$,
and a greater portion of the velocity signature is fluctuating near
$V = 0$.
The average velocity for the $\omega/2\pi=7.5 \times 10^{-5}$ system
is $\langle V\rangle=0.12$, while
for the $\omega/2\pi=5\times 10^{-6}$ system it is
$\langle V\rangle=0.02$,
or six times lower, indicating the strong frequency dependence of the
ratchet efficiency.
In Fig.~\ref{fig:14}(d), we plot the particle configurations
during the $B = 0.0$ portion of the cycle
for the $\omega/2\pi = 5\times 10^{-6}$ system,
showing the formation of a more uniform stripe-like pattern.

\section{Substrate Periodicity}

We next fix the particle density to $\rho = 0.52$ and
continue to sweep $B$ up to a maximum value of
$B=4.2$, but vary the number of substrate minima $N_{p}$, which changes
the substrate lattice constant $a$.
For clarity, in this section we measure
the non-normalized average velocity $\langle V \rangle$.
In the previous sections, we considered samples with
$N_p = 35$ in which the bubble diameters were generally larger
than the substrate lattice spacing. Here, that is not always the case.

\begin{figure}
\includegraphics[width=\columnwidth]{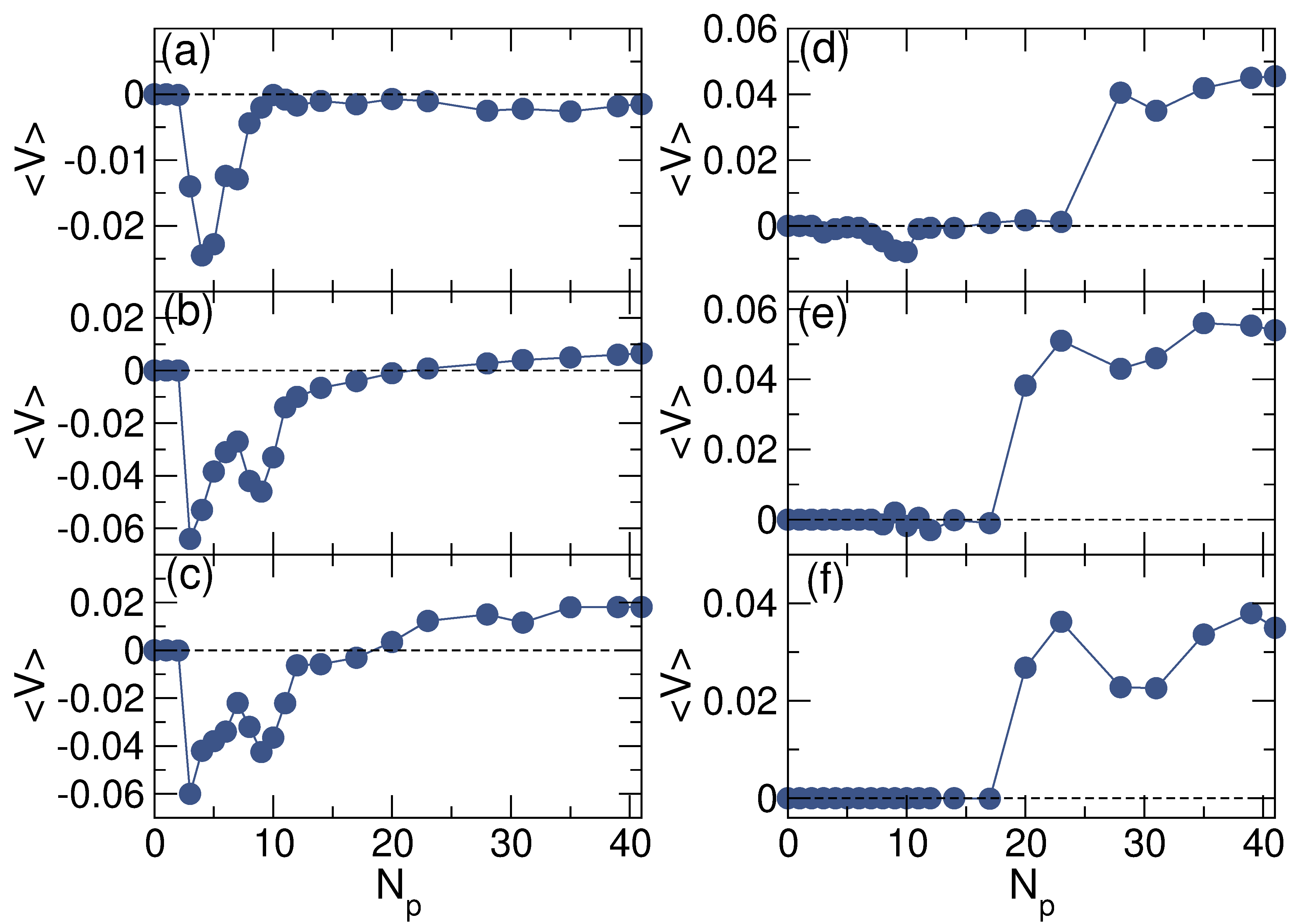}
\caption{$\langle V \rangle$ vs the number of substrate
minima $N_{p}$ in systems with $\rho = 0.52$ and $\omega/2\pi=2.5\times 10^{-5}$.
(a) $A_{p} = 0.125$ showing a negative ratchet.
(b) $A_{p} = 0.5$,
where there is a crossover
from a negative ratchet effect to a positive ratchet
effect for $N_{p} > 2.0$.
(c) $A_{p} = 0.75$.
(d) $A_{p} = 1.5$,
where the negative ratchet effect at lower $N_{p}$ is
of decreased magnitude.
(e) $A_{p} = 2.0$, where only a positive ratchet effect occurs.
(f) $A_{p} = 3.2$, where only a positive ratchet effect occurs.
}
\label{fig:16}
\end{figure}

\begin{figure}
\includegraphics[width=\columnwidth]{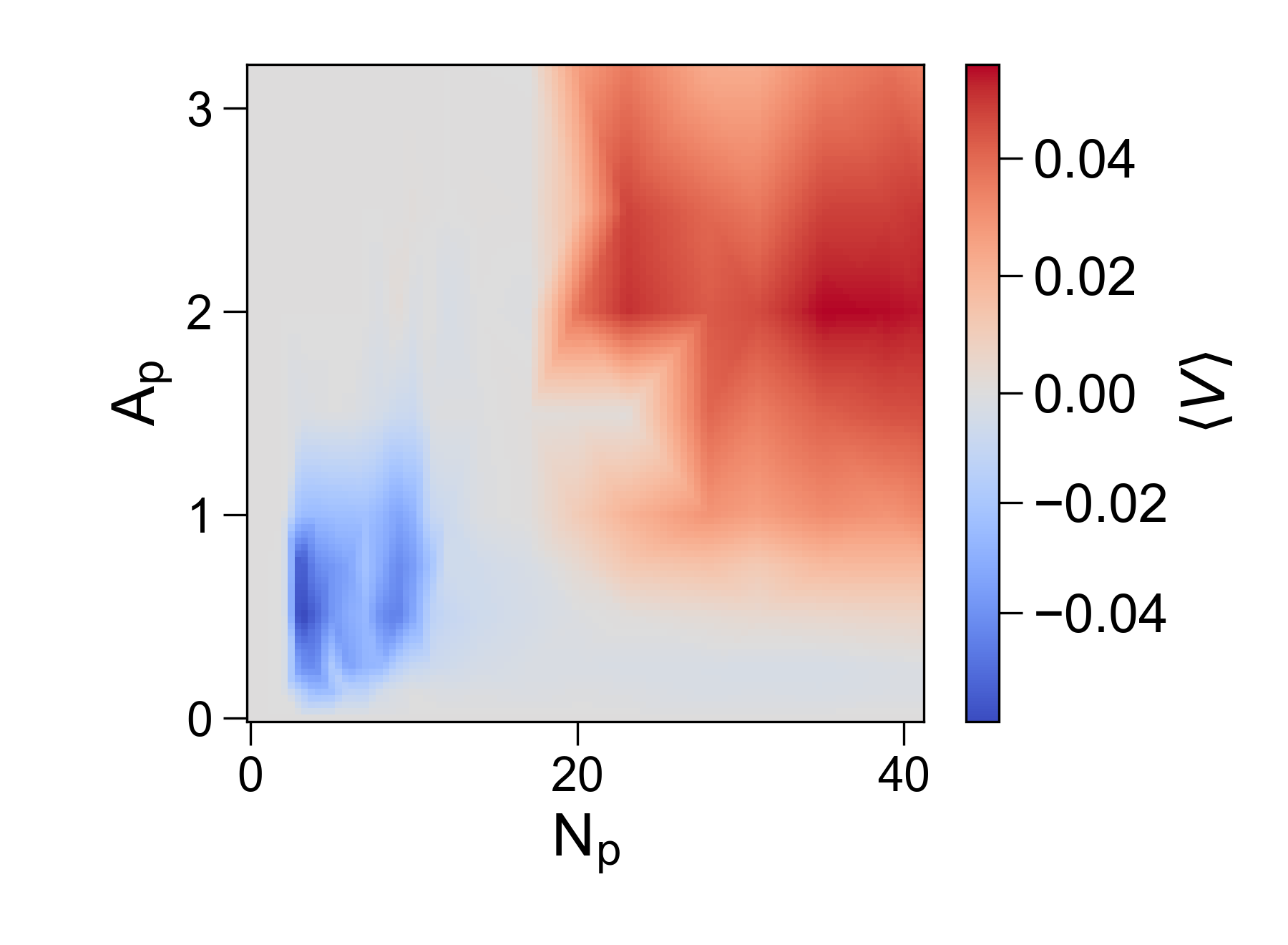}
\caption{Heatmap of $\langle V\rangle$ as a function of
$A_p$ vs $N_p$ for the system
from Fig.~\ref{fig:16} with $\rho=0.52$ and $\omega/2\pi=2.5 \times 10^{-5}$
showing both negative and
positive ratchet effect regimes.
}
\label{fig:17}
\end{figure}

In Fig.~\ref{fig:16},
we plot $\langle V \rangle$ versus $N_{p}$
for varied pinning strength $A_p$.
For $N_p < 3.0$, the ratchet effect is absent or strongly reduced,
since the spacing between the substrate maxima
is larger than the motion of the particles.
At $A_p=0.125$ in Fig.~\ref{fig:16}(a), as shown in the previous
sections there is a weak negative ratchet effect when
$N_p = 35$. This negative
ratchet effect becomes more pronounced
as $N_{p}$ decreases below $N_p=35$, and reaches its
maximum efficiency at $N_p = 4$.
In Fig.~\ref{fig:16}(b), at $A_p = 0.5$,
there is a transition
from a negative ratchet effect
for $N_p < 22$ to
a positive ratchet effect for $N_p > 22$.
The negative ratchet effect is larger than in the $A_p=0.125$ system,
and we find the greatest
ratchet efficiency near $N_p = 4$.
At $A_p=0.75$ in Fig.~\ref{fig:16}(c),
the positive ratchet effect is stronger and
occurs when $N_{p} > 20$.
We also find a local minimum in the efficiency of the
negative ratchet effect near $N_{p} = 7$.
Figure~\ref{fig:16}(d) shows that at $A_p = 1.5$,
the negative ratchet effect at low $N_p$ is reduced in magnitude,
while the positive ratchet effect that appears for higher $N_{p}$
has become stronger.
In Fig.~\ref{fig:16}(e) at $A_p=2.0$ and Fig.~\ref{fig:16}(f) at
$A_p=3.2$,
there is only a positive ratchet effect,
which appears when  $N_p > 20$.
Figure~\ref{fig:17} shows a heatmap of $\langle V\rangle$ as a
function of $A_p$ versus $N_p$ for the system from Fig.~\ref{fig:16}.
A negative ratchet effect appears
for $A_p < 0.75$, and reaches its greatest efficiency for
$2 < N_{p} < 11$,
while when $A_p > 0.75$ and $N_{p} > 20$, there
is a positive ratchet effect.

\begin{figure}
\includegraphics[width=\columnwidth]{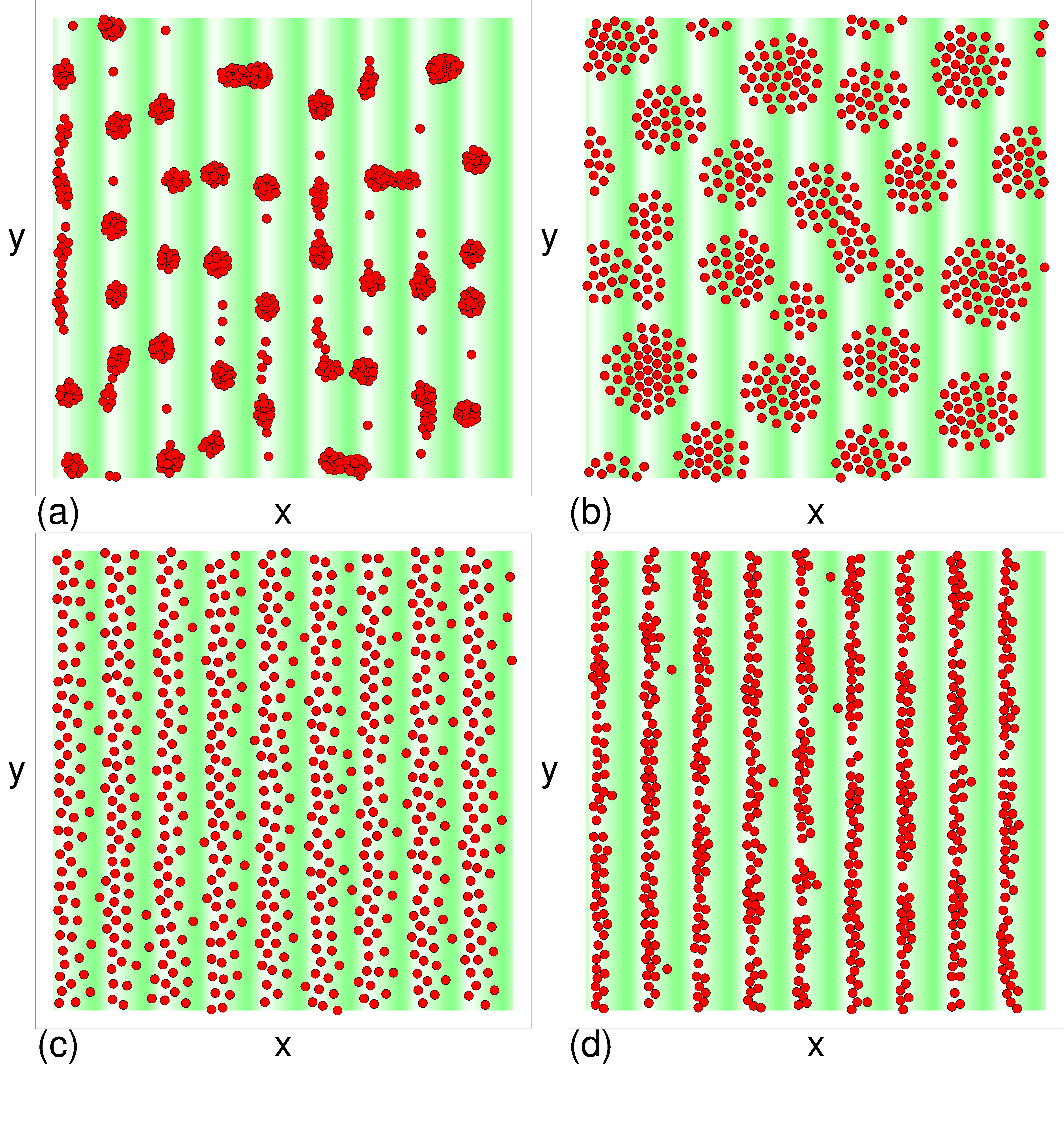}
\caption{Image of particle positions (red) and the underlying
asymmetric potential (green) during different portions of the cycle
for the sample from Fig.~\ref{fig:16}(b)
with $A_p=0.5$, $\rho=0.52$, $\omega/2\pi=2.5\times 10^{-5}$,
and $N_p=9$, where there is a negative
ratchet effect.  
(a) The $B = 2.4$ state after a single cycle has been completed.
(b) Bubble expansion at $B=2.6$ during the decreasing $B$ portion of the cycle.
(c) Stripes of relatively uniform density
at $B = 0.0$.
(d) Close to the end of the increasing $B$
portion of the cycle near $B = 2.4$, just before the bubbles collapse.
}
\label{fig:18}
\end{figure}

\begin{figure}
\includegraphics[width=\columnwidth]{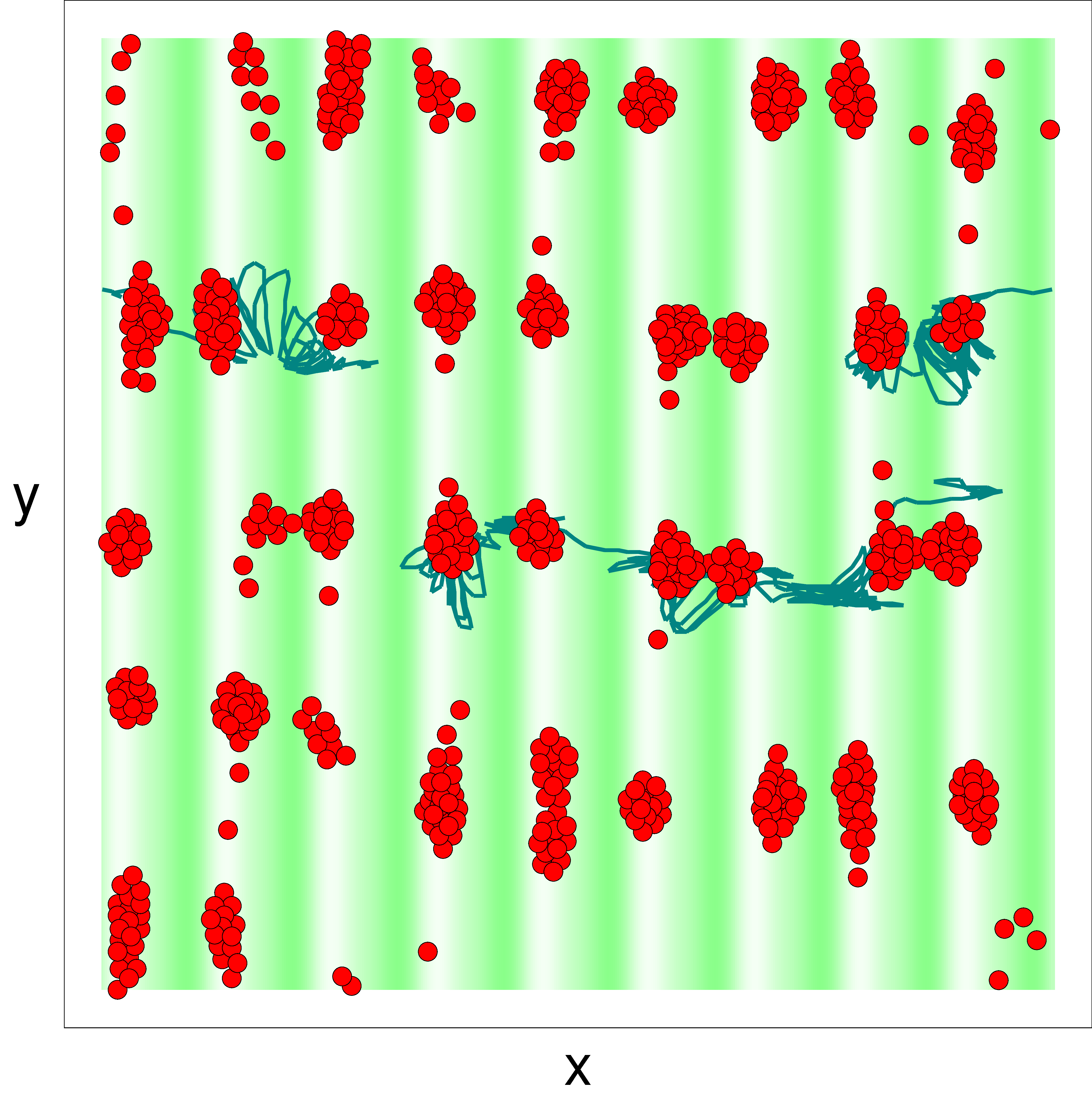}
\caption{Image of particle positions (red) and the underlying asymmetric
potential (green) along with trajectories (lines) of two particles during
20 cycles in the $A_p=0.5$, $\rho=0.5$,
$\omega/2\pi=2.5\times 10^{-5}$, and $N_p=9$ sample from
Fig.~\ref{fig:18} where there is a negative ratchet effect.
}
\label{fig:19}
\end{figure}

To better understand the difference between the negative and
positive ratchet effects,
in Fig.~\ref{fig:18}, we illustrate
the particle positions during one cycle for the system
from Fig.~\ref{fig:16}(b) with
$A_p = 0.5$ and $N_p = 9$, where there is a negative ratchet effect.
Figure~\ref{fig:18}(a) shows the $B=2.4$ configuration after a single cycle
has completed, where compact clusters are present.
As $B$ decreases, the bubbles begin to expand, as shown at $B=2.6$ in
Fig.~\ref{fig:18}(b).
At $B=0.0$ in Fig.~\ref{fig:18}(c), stripes of relatively uniform
density form. When $B$ increases again, the system remains in a stripe
configuration during nearly the entire cycle, as shown in
Fig.~\ref{fig:18}(d) where $B$ is only slightly below $B=2.4$ but the bubbles
have not yet collapsed. Once $B$ reaches $B=2.4$, the bubbles collapse
and the system returns to a state
similar to that shown in Fig.~\ref{fig:18}(a).
The asymmetry between the presence of bubbles during the decreasing $B$ portion
of the cycle and the presence of stripes during the increasing $B$ portion
of the cycle is responsible for producing the negative ratchet effect.
To more clearly show that a ratchet effect is occurring,
in Fig.~\ref{fig:19} we plot
the particle positions and the trajectories of two particles for
the system from Fig.~\ref{fig:18},
demonstrating that the particles are translating in the negative
$x$ direction.

\begin{figure}
\includegraphics[width=\columnwidth]{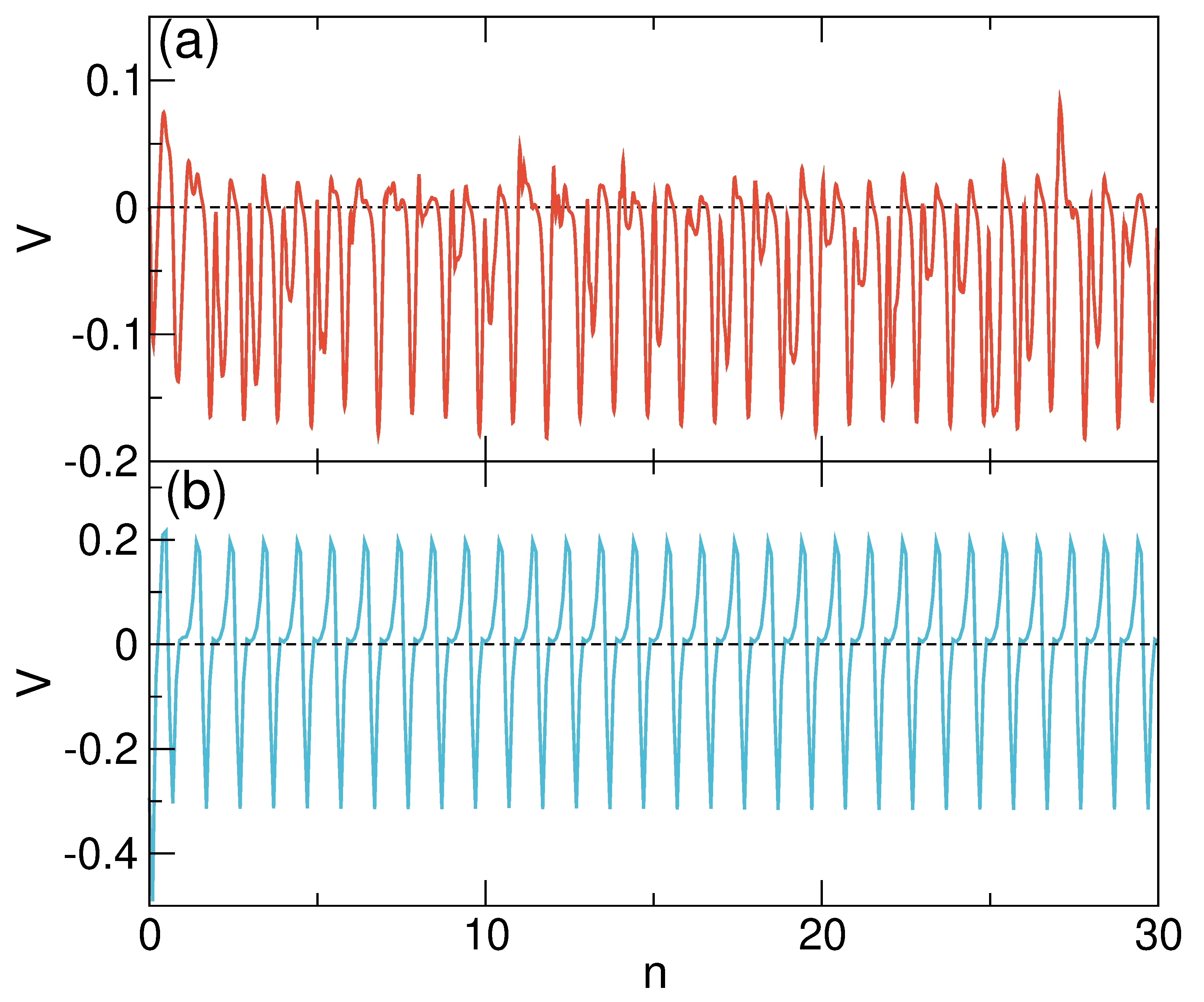}
\caption{The instantaneous average velocity $V$
vs time in cycle numbers $n$ for the system in Fig.~\ref{fig:16}
with $\rho=0.52$ and $\omega/2\pi=2.5\times 10^{-5}$.
(a) At $N_p = 9$ and $A_p = 0.5$, there is a negative ratchet effect.
(b) At $N_p = 4$ and $A_p = 2.0$, there is no ratchet effect.}
\label{fig:20}
\end{figure}

\begin{figure}
\includegraphics[width=\columnwidth]{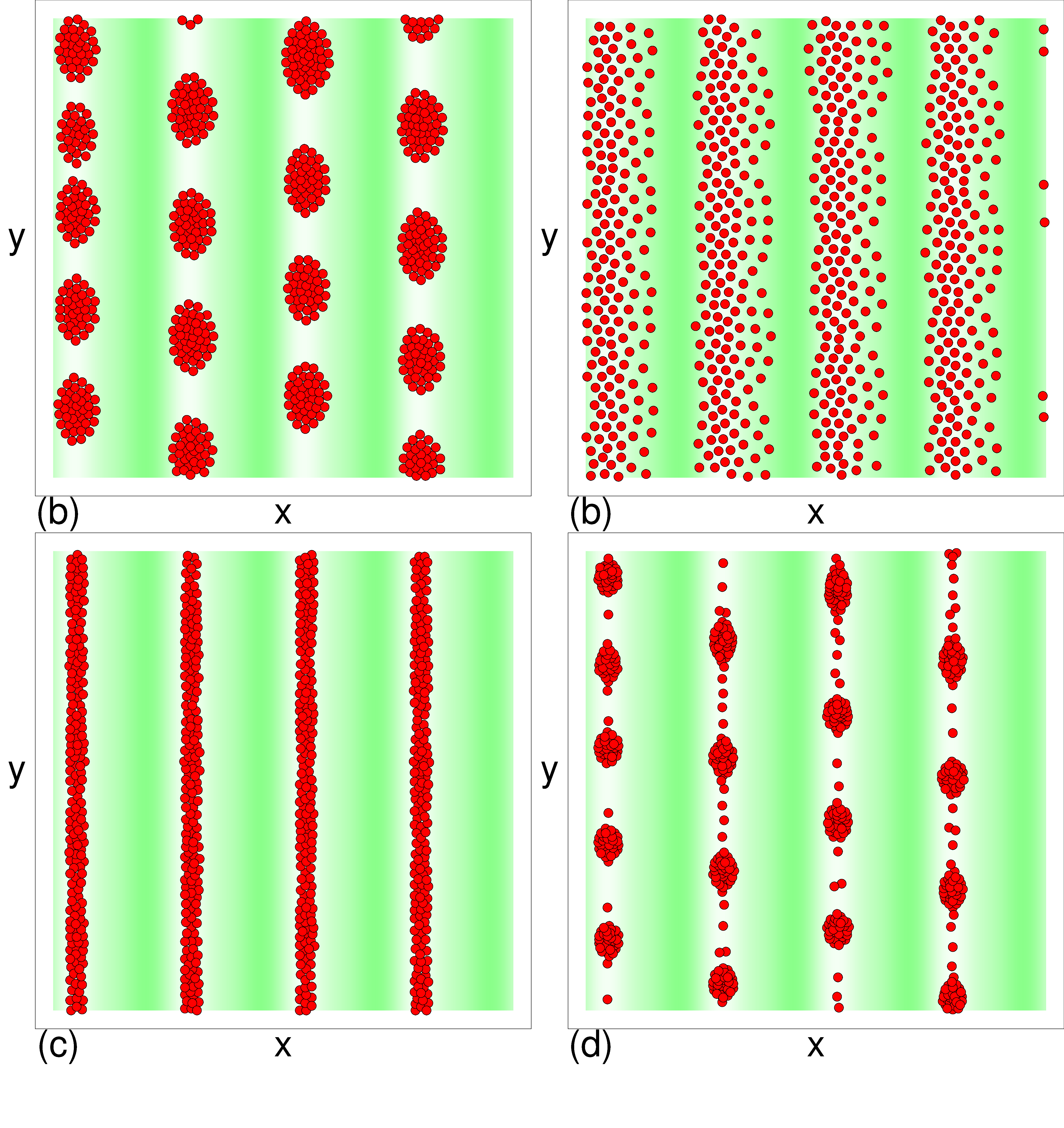}
\caption{Image of particle positions (red) and the underlying
asymmetric potential (green) during different portions of the cycle
for the system from Fig.~\ref{fig:20}(b) with $A_p = 2.0$, $N_p = 4.0$,
$\omega/2\pi=2.5\times 10^{-5}$,
and $\rho = 0.52$.
There is no ratchet effect, and the particles undergo
oscillations within each well without traveling from one well to another.
(a) The $B=4.0$ state in the decreasing $B$ portion of the cycle after the
bubble expansion has begun.
(b) $B=0.0$, where the stripes each exhibit a density gradient from the
asymmetric substrate, but there is no hopping of particles from well to well.
(c) The $B=2.0$ state during the increasing $B$ portion of the cycle when
the stripes are becoming more compact.
(d) The $B=3.8$ state in the increasing $B$ portion of the cycle, where
the bubbles are beginning to reform.}
\label{fig:21}
\end{figure}

When both $A_p$ and $N_p$ have low values,
there are particles moving in both the negative and positive $x$ directions
during the bubble expansion phase that occurs when $B$ is decreasing.
Since each bubble was centered on a substrate minimum at $B=2.4$,
the particles moving in the negative $x$ direction translate by a
smaller distance before reaching the edge of the pinning barrier
compared to particles moving in the positive $x$ direction, due to the
substrate asymmetry.
When $A_p$ is small, the pinning barrier is too low to stop
all of the particles from hopping across it, while when $A_p$ is larger,
some of the particles traveling in the positive $x$ direction will not have
enough time during the bubble expansion phase to reach the barrier edge and
jump across it.
During the deceasing $B$
or bubble expansion portion of the cycle, more particles are able to
hop over the barrier in the negative $x$ direction than in the positive $x$
direction due to the different distances to the barrier edge.
In contrast, during the increasing $B$ or bubble reformation portion of the
cycle, only some of the particles are able to jump back across the barrier
and rejoin their original bubble, while other particles have gone too far
across the barrier and instead become part of a bubble that is on the
other side of the barrier. These particles undergo a net translation in the
negative $x$ direction, giving a relatively weak negative ratchet effect.
As the number of substrate minima decreases, the asymmetry in the distances
that particles must travel in the positive and negative $x$ directions to
reach the barrier edge also increases, so that relatively fewer particles
are able to cross the barrier in the easy or positive $x$ direction compared
to the hard or negative $x$ direction, 
leading to the enhancement of the negative ratchet
effect at lower $N_p$ shown in
Fig.~\ref{fig:16}.
When $A_p$ is made larger, the particles have a more difficult time
jumping over the barrier edge in the hard direction,
and the negative ratchet effect is diminished or extinguished.
At higher $N_p$ and high $A_p$,
the particles are prevented by the pinning strength
from jumping over the barrier in the hard or negative $x$ direction,
but since the distance that must be traveled to jump over the
barrier in the easy or positive $x$ direction has decreased due to the
smaller substrate lattice constant, more particles are able to make this
jump successfully, resulting in the emergence of
a positive ratchet effect. For the parameters shown in Fig.~\ref{fig:16},
this transition to positive ratchet motion occurs near $N_p = 20$.
When $A_p$ is too large, there is not enough time for
the particles to jump over either barrier edge during the
bubble expansion portion of the cycle, and the ratchet effect is
completely lost.

\begin{figure}
\includegraphics[width=\columnwidth]{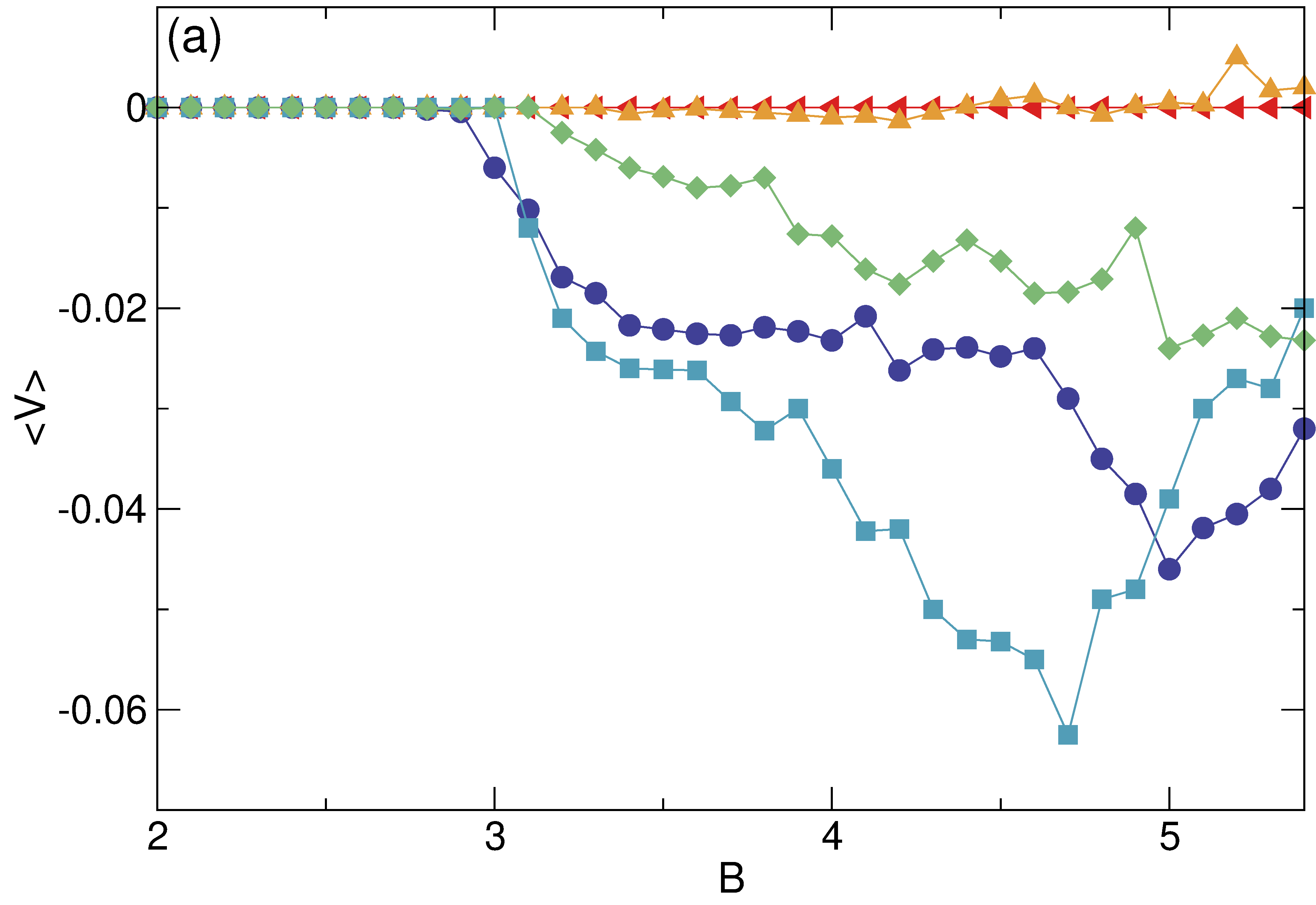}
\includegraphics[width=\columnwidth]{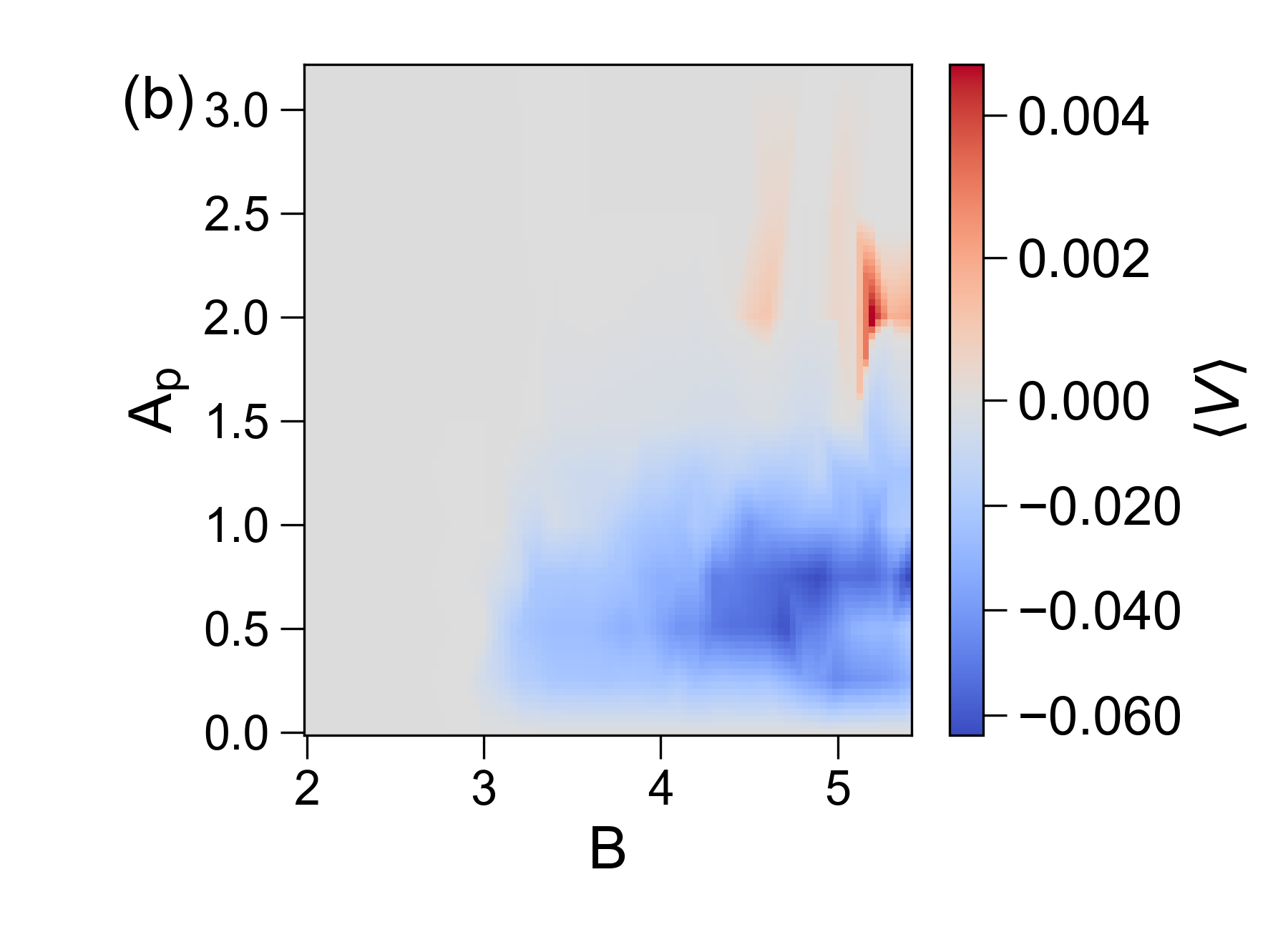}
\caption{(a) Average ratchet velocity $\langle V\rangle$
vs the maximum value of
$B$ for a system with $N_p = 8$, $\omega/2\pi=2.5\times 10^{-5}$,
and $\rho = 0.52$ at
$A_{p} = 0.25$ (dark blue circles),
$A_p = 0.5$ (light blue squares),
$1.25$ (green diamonds),  
$2.0$ (orange up triangles),
and $3.2$ (red left triangles).
(b) Heatmap of $\langle V\rangle$ as a
function of $A_p$ vs the maximum value of $B$ for the same system.
}
\label{fig:22}
\end{figure}

Figure~\ref{fig:20}(a) shows the time series
of the instantaneous velocity $V$
for the system from Fig.~\ref{fig:18} with $\rho=0.52$,
$N_p = 9$, and $A_p = 0.5$. The velocities are clearly biased in the
negative $x$ direction.
The maximally negative values of $V$ occur at
the $B = 0$ portion of the cycle,
where some particles jump over the barrier in the negative $x$ direction
and a smaller number of particles are able to jump
over the barrier in the positive $x$ direction.
In Fig.~\ref{fig:20}(b), we show the system from Fig.~\ref{fig:16}(e) at
$\rho=0.52$, $A_p = 2.0$ and $N_p = 4$, where there is no
ratchet effect.
Here, the velocity oscillates
asymmetrically around zero and has a zero average value.
Figure~\ref{fig:21} shows the particle configurations during one cycle
for the $A_p=2.0$ and $N_p=4$ sample from Fig.~\ref{fig:20}(b).
The expanding bubbles appear at $B=4.0$ in Fig.~\ref{fig:21}(a).
At $B=0.0$ in Fig.~\ref{fig:21}(b), the particles have spread into
stripes with a density modulation induced by the asymmetric substrate.
The contraction of the stripes back into bubbles is shown
in Fig.~\ref{fig:21}(c) at $B=2.0$ and in Fig.~\ref{fig:21}(d) at
$B=3.8$. In each case, there is never any hopping of particles from one
minimum to an adjacent minimum because the substrate lattice constant
is larger than the maximum distance the particles are able to travel
during a single cycle, and so no ratchet effect can arise.

\begin{figure}
\includegraphics[width=\columnwidth]{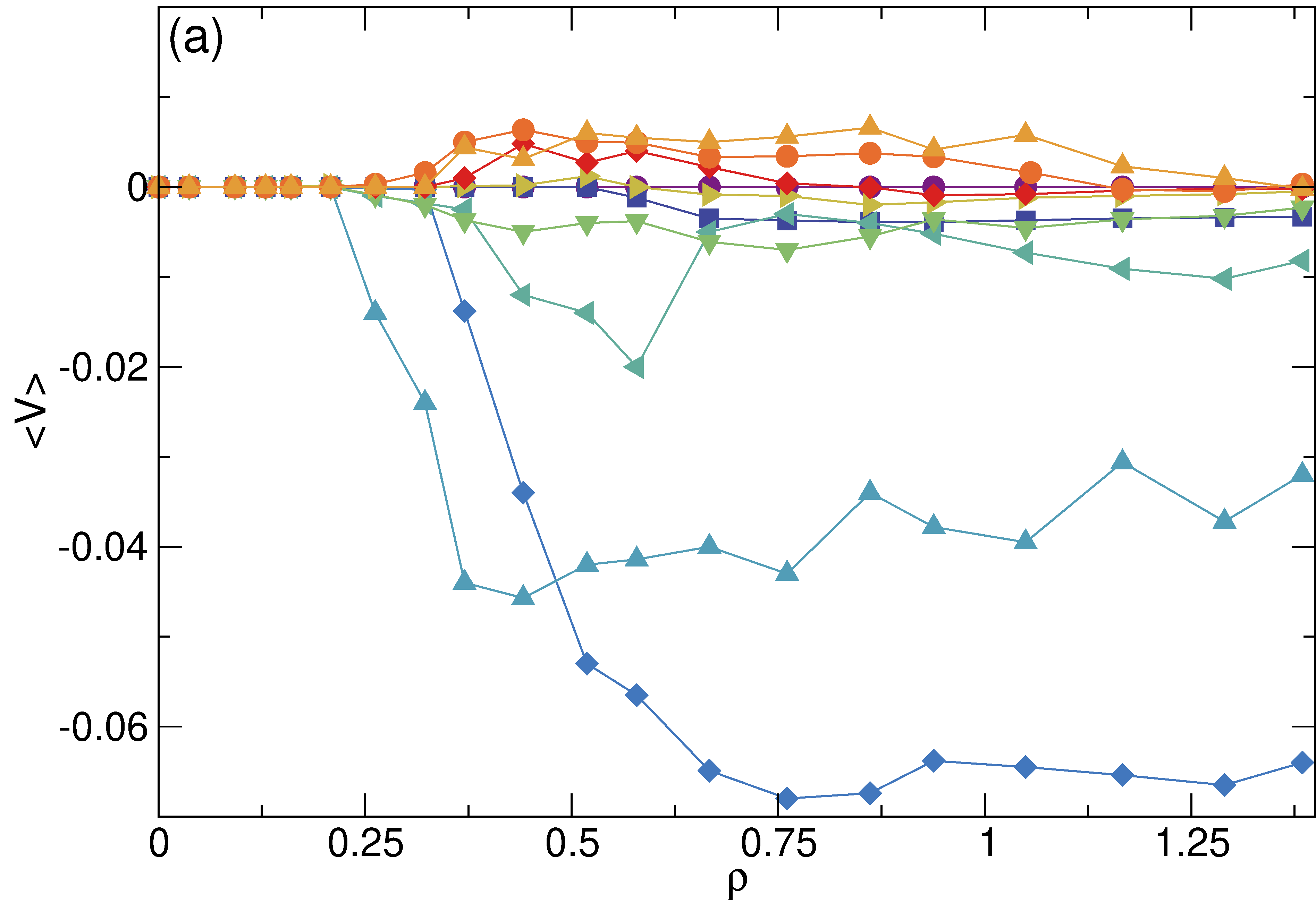}
\includegraphics[width=\columnwidth]{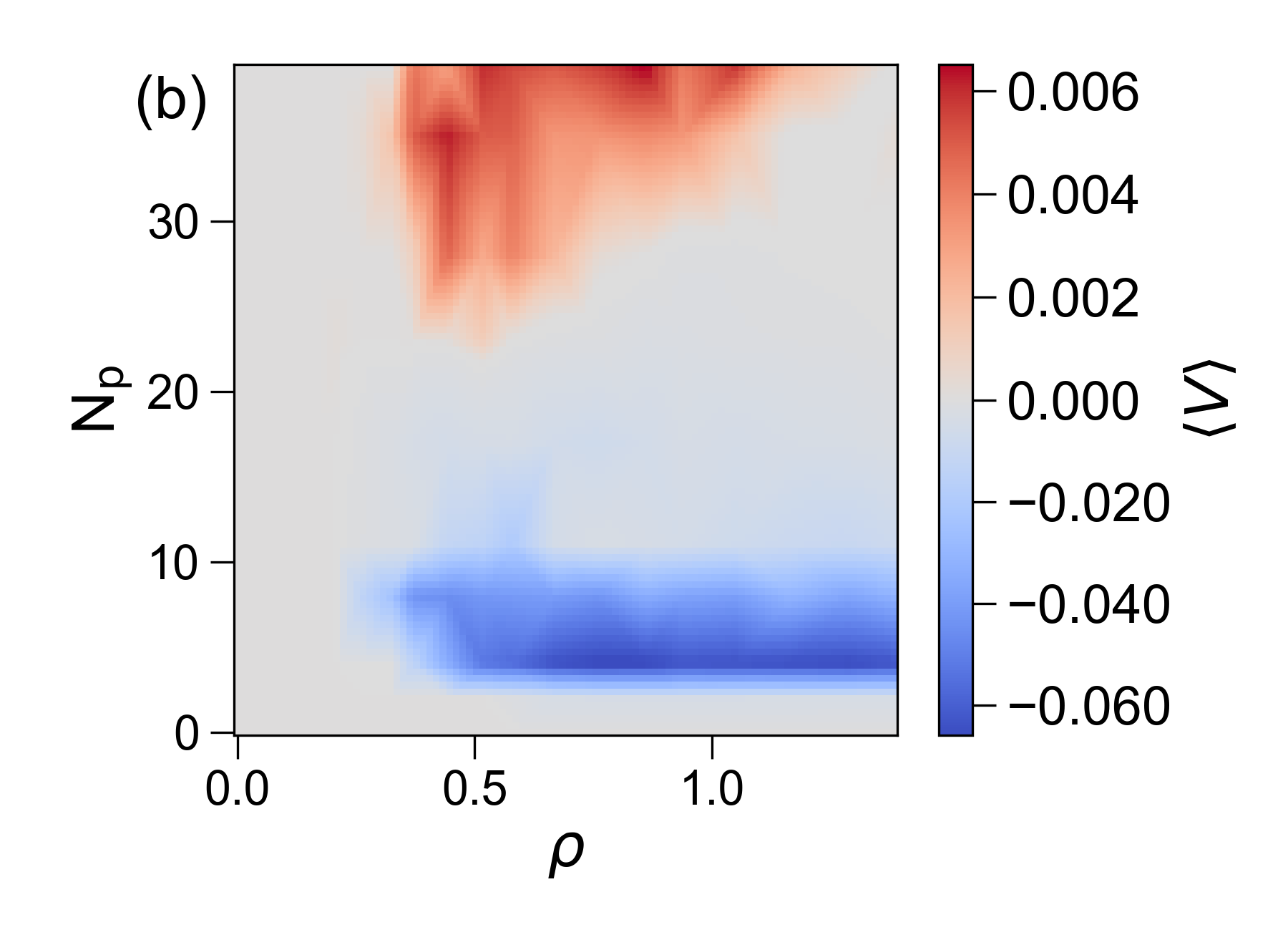}
\caption{(a) $\langle V \rangle$ vs $\rho$
for samples with $A_p=0.5$, $\omega/2\pi=2.5\times 10^{-5}$, and
a fixed maximum value of $B = 4.2$ at
$N_p = 0$ (violet circles),
2 (dark blue squares),
4 (medium blue diamonds),
8 (light blue up triangles),
11 (dark green left triangles),
17 (medium green down triangles),
23 (light green right triangles),
28 (light orange up triangles),
35 (dark orange circles),
and 39 (red diamonds).
(b) The corresponding heatmap of $\langle V\rangle$
as a function of $N_p$ vs $\rho$
showing regions of negative and positive ratchet effects.
}
\label{fig:23}
\end{figure}

In Fig.~\ref{fig:22}(a),
we show $\langle V \rangle$ versus the maximum
value of $B$
for a system with $N_p = 8$, $\omega/2\pi=2.5\times 10^{-5}$,
and $\rho = 0.52$ at $A_p = 0.25$, 0.5, 1.25, 2.0, and 3.2.
Here, there is a critical maximum $B$ value
required for the ratchet effect to occur,
and the largest negative ratchet effects
appear for $A_p = 0.5$ at $B = 4.7$.
As $A_p$ increases, the efficiency of the negative ratchet increases,
but for sufficiently large $A_p$, the ratchet efficiency decreases
again as the
particles become confined by the substrate and are unable to hop between
adjacent substrate minima.
Figure~\ref{fig:22}(b) shows a heatmap of $\langle V\rangle$ as a function
of $A_p$ versus the maximum value of $B$
for the system from Fig.~\ref{fig:22}(a)
including data from additional parameters, indicating the region in which
the negative ratchet effect occurs.
There are also small regions where there is a positive ratchet effect
for $B = 5$ and $A_p > 1.5$.

\begin{figure}
\includegraphics[width=\columnwidth]{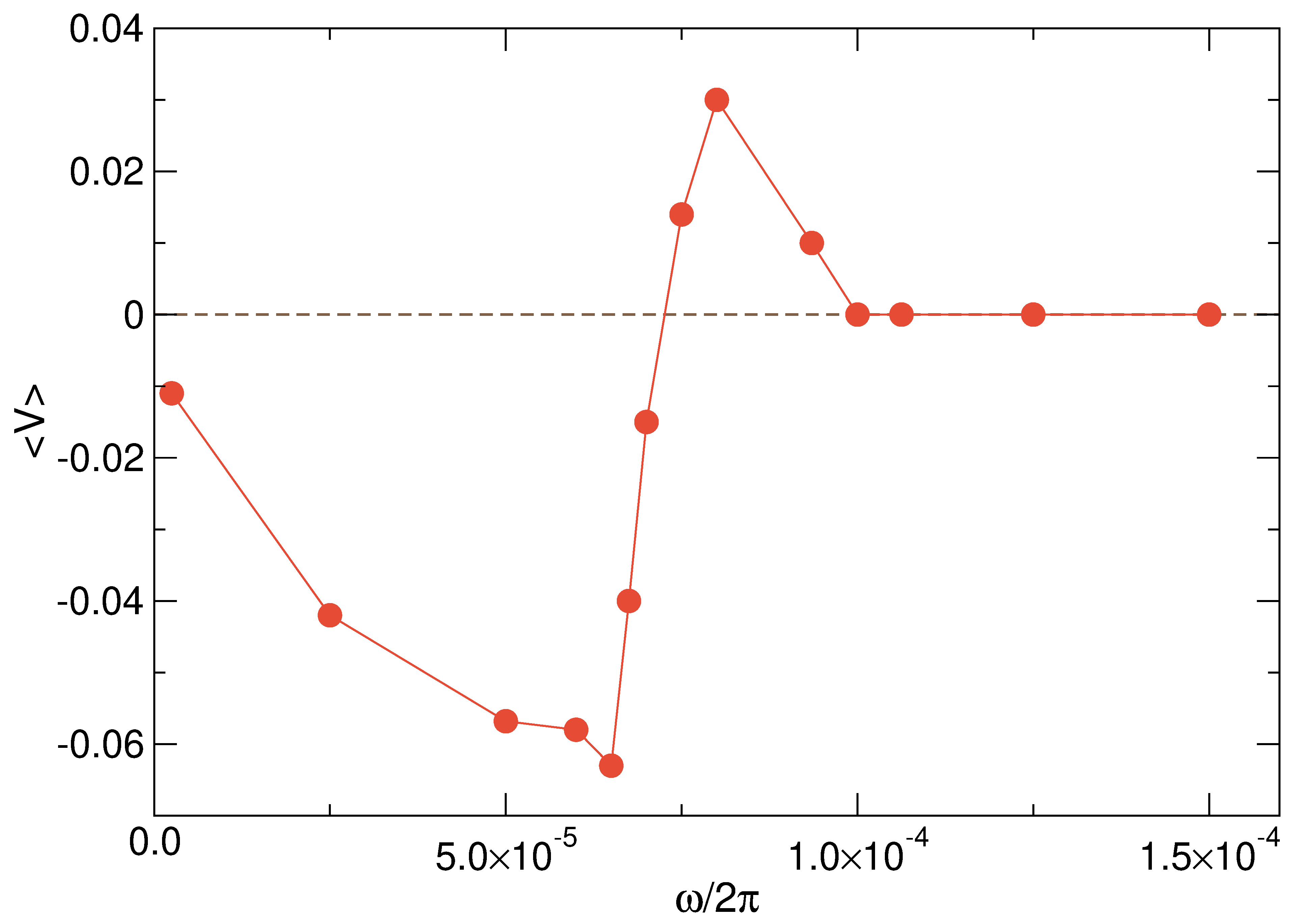}
\caption{ $\langle V\rangle$ vs $\omega/2\pi$ for
a system with $\rho = 0.52$, $A_p = 0.5$,
maximum $B$ of $B=4.2$,
and $N_p  = 8$ showing
that there is a reversal from a negative to a positive
ratchet effect as the frequency increases.
}
\label{fig:24}
\end{figure}

Figure~\ref{fig:23}(a)
is a plot of $\langle V \rangle$ versus
$\rho$ for a system with
$A_p=0.5$, $\omega/2\pi=2.5\times 10^{-5}$, and a
fixed maximum value of $B = 4.2$ at
$N_p = 0$, 2, 4, 8, 11, 17, 23, 28, 35, and 39.
For $\rho < 0.25$, there is no ratchet effect.
The largest negative ratchet effect appears
at $N_p = 4$, and the ratchet efficiency decreases with increasing $N_p$
until there is a crossover to
a positive ratchet when $N_p > 20$.
For this value of $A_p$,
the positive ratchet effect is weak,
but if $A_p$ is made larger,
the efficiency of the negative ratchet effect decreases
while the efficiency of the positive ratchet effect increases.
In Fig.~\ref{fig:23}(b), we
plot a heatmap of $\langle V\rangle$ as a function of
$N_p$ versus $\rho$ for the same system,
showing that there is a strong negative ratchet for
$N_p < 10$ and that a ratchet reversal occurs for $N_p > 10$.

We have also examined the frequency dependence
of the ratchet efficiency in the negative ratchet regime.
In Fig.~\ref{fig:24}, we plot $\langle V \rangle$ versus
$\omega/2\pi$ for a system with
$\rho = 0.52$, $A_p = 0.5$, $N_p = 8$, and a maximum $B$ value of
$B=4.2$.
The efficiency of the negative ratchet is maximized at
$\omega/2\pi=6.5\times 10^{-5}$, and as the frequency is
increased further, a reversal occurs to a positive ratchet effect.
All ratcheting motion dies away once the frequency becomes too high
since the particles are unable to escape from the individual wells during
a single oscillation cycle when the cycle is very short.
This result indicates
that ratchet reversals can also occur as a function of frequency.

\section{Discussion}

In reviewing the literature, the closest effect
we found to the ratchet system that we propose in this work
is a single skyrmion bubble moving in an
asymmetric channel under an oscillating magnetic field
\cite{Migita20}.
The size of the skyrmion changes as the magnetic field changes,
resulting in a ratcheting motion of the skyrmion
along the easy direction of the substrate potential.
This behavior is similar to what we observe if,
for our bubble phase, the system is viewed as containing a bubble
of varying size.
An important difference between our system and the skyrmion bubble
is that for the SALR particles, a strong breakup of the bubbles
and mixing of the particles occurs during the ratcheting motion.
This suggests that ratchet effects can arise
not only due to changes in the size of an object, but also from
changes in its shape and/or morphology.
In our system, the motion is partially disordered due to the
breakup of the bubble during the oscillation cycle.
Under conditions where the bubbles do not break apart during the oscillation,
it is possible that the bubbles would show a more orderly ratcheting
motion.
Adding thermal fluctuations to our system 
would likely reduce the efficiency of the ratchet motion,
but it may also 
extend the range of parameters over which a finite ratchet effect is
present,
such as by making ratcheting possible
for higher substrate strengths or
for lower particle densities.

\section{Summary} 

We have proposed a new collective ratchet effect in which
directed motion can occur for a pattern-forming system on an
asymmetric substrate when the properties of the particle-particle
interaction potential are periodically oscillated.
Most past observations of ratchet effects have been performed
for particles on asymmetric substrates subjected to ac driving or to
flashing of the substrate.
We specifically consider a pattern-forming system of
particles with competing short-range attractive
and long-range repulsive (SALR) interactions.
In the absence of a substrate, bubble, stripe, and crystal states can form
depending on the ratio of the attraction to the repulsion.
We place the SALR particles on a periodic,
quasi-one-dimensional asymmetric substrate
and cycle the ratio of the attractive term to the repulsive term
so that the system oscillates
between a bubble phase for strong attraction
and a uniform crystal for pure repulsion.
We observe a ratchet effect in which the particles have a net translation
along the easy substrate asymmetry direction as a result of the
spatially asymmetric
spreading out of the bubbles during the portion of the cycle in which the
attractive term is decreased. More of the particles from each bubble move
in the easy substrate direction than in the hard substrate direction.
The ratchet efficiency is strongly sensitive to the oscillation frequency,
and is reduced at low frequencies because the gradient in particle density
produced by the asymmetrically spreading bubbles has enough time to relax
toward a uniform nonratcheting state.
There is an optimal frequency at which the gradient is best stabilized
during the zero-attraction portion of the cycle, and
this maximizes the efficiency of
the ratchet motion.
At high frequencies, the particles remain trapped
in individual oscillating bubbles and are unable
to travel from one substrate minimum
to the next, so the ratchet effect is lost.
We show that there is an optimal ratchet efficiency
as a function of particle density, substrate strength,
and substrate lattice spacing.
At low particle densities, the bubbles cannot form
and the ratchet effect is lost.
If the maximum attractive force reached
during the oscillation cycle is too small,
bubbles do not form and the ratchet effect is absent.
When the substrate is weak or the substrate lattice spacing is small,
we find that 
ratchet reversals can occur in which there is
a transition from a negative to a positive ratchet
effect as a function of substrate strength, substrate lattice
spacing, and frequency.
There are numerous systems with competing interactions that exhibit
patterns similar to what we consider here,
and our results suggest that such systems
could exhibit a ratchet effect
if the pattern is changed periodically
when the system is coupled to an asymmetric substrate.
The ratchet effect we describe
could be realized using colloidal particles
with competing interactions that are generated by a magnetic field,
where the field could be oscillated, or in
certain bubble-forming systems where
the pattern changes shape as a function of an oscillatory field.

\begin{acknowledgements}
We gratefully acknowledge the support of the U.S. Department of
Energy through the LANL/LDRD program for this work.
This work was supported by the US Department of Energy through
the Los Alamos National Laboratory.  Los Alamos National Laboratory is
operated by Triad National Security, LLC, for the National Nuclear Security
Administration of the U. S. Department of Energy (Contract No. 892333218NCA000001).
\end{acknowledgements}

\bibliography{mybib}

\end{document}